\documentclass[a4paper,aps,pre,amssymb,twocolumn,superscriptaddress,showpacs]{revtex4}
\usepackage{graphicx}
\usepackage{color}
\usepackage{amsmath}

\def\e{\mathrm{e}}
\def\i{\mathrm{i}}
\def\d{\mathrm{d}}

\begin{document}

\title{Spin Glass Model of Operational Risk}

 \author{M. Bardoscia} \affiliation{Dipartimento di Fisica,
 Universit\`a di Bari,
        I-70126  Bari, Italy}
 \affiliation{INFN, Sezione di Bari, I-70126 Bari, Italy}
 
 \author{P. Facchi}
 \affiliation{Dipartimento di Matematica, Universit\`a di Bari,
         I-70125  Bari, Italy}
 \affiliation{INFN, Sezione di Bari, I-70126 Bari, Italy}
 
 \author{S. Pascazio} \affiliation{Dipartimento di Fisica,
 Universit\`a di Bari,
        I-70126  Bari, Italy}
 \affiliation{INFN, Sezione di Bari, I-70126 Bari, Italy}
 
 \author{A. Trullo} \affiliation{Dipartimento di Fisica,
 Universit\`a di Bari,
        I-70126  Bari, Italy}
 \affiliation{INFN, Sezione di Bari, I-70126 Bari, Italy}
\date{\today}

\begin{abstract}
We analyze operational risk in terms of a spin glass model. Several regimes are investigated, as a functions of the parameters that characterize the dynamics. The system is found to be robust against variations of these parameters. 
We unveil the presence of limit cycles and scrutinize the features of the asymptotic state.
\end{abstract}


\maketitle

\section{Introduction}

The physics of complex systems has proven to be a fertile arena of investigation in very diverse areas of natural sciences. 
The versatility of the techniques of analysis and the wide applicability of the main underlying ideas make this subject interdisciplinary: there are applications in physics, mathematics, statistics, cybernetics, chemistry, biology, as well as economics and social sciences, this list being far from complete. 

The guiding principles in the study of complex systems are philosophically intriguing, and lie in the idea that one can learn simple lessons from complexity \cite{GK}. This requires the definition of an appropriate level of description at which the main features of the system investigated can be described in terms of a given number of suitable variables. The very definition of these variables and their interactions design a convenient ``minimal model," whose dynamics, equilibrium features and correlations can be investigated. 

Spin glasses play an important role in this context \cite{parisi, dotsenko}. 
A spin glass can be thought of as an ensemble of interacting spins, whose bonds are randomly distributed. The magnetic ordering that emerges resembles the positional ordering of a physical glass and is characterized by frustrated interactions. Like all complex systems, a spin glass is composed of interconnected parts (spins) and as a whole exhibits properties that are far from being obvious if one looks at the properties of the individual parts.

In this article we shall apply a spin glass model to study a financial system \cite{mandelbrot,MantegnaStanley}. 
We shall seek an appropriate level of description at which capital losses can be schematized as interacting dichotomic variables (spins).
We shall focus in particular on ``Operational Risk" (OR) \cite{cruz, king}, defined by the The New Basel Capital Accord (Basel II) \cite{basel}
as ``the risk of [money] loss resulting from inadequate or failed internal processes, people and systems or from external events''. OR management \cite{embrechts} is a subcategory of risk management, that branch of economy whose purpose is the identification of the possible sources of money losses and eventually the definition of a strategy for avoiding them. According to the regulations set by the ``Basic Indicator Approach" proposed in Basel II, banks have to set aside 15\% of their total capital, in order to face operational losses: since this is a quenched capital, banks share a keen interest in order to keep this percentage as small as possible. 

Basel II also proposes a slightly refined approach (the ``Standardized Approach'') that defines eight different ``business lines" \footnote{The eight business lines are \cite{basel}:
Corporate Finance, Trading \& Sales, Retail Banking, Commercial Banking, Payment \& Settlement, Agency Services, Asset Management and Retail Brokerage.}
on which losses can occur, by associating to each of them a constant risk coefficient that can vary between 12\% and 18\%, so that the total amount of money to be stored by a bank is obtained by a weighted sum. The Basel accord gives to each bank the possibility to propose an internally developed ``Advanced Measurement Approach'' to estimate the capital to set aside. Let us point out that, while the Basic Indicator Approach and the Standardized Approach are aimed only to the estimation of this capital, the Advanced Measurement Approaches are particularly attractive for a bank because they may be oriented to the management of OR. Every Advanced Measurement Approach is required to classify the operational losses in the 8 afore-mentioned business lines and in 7 ``event types''
\footnote{The seven event types are \cite{basel}: Internal Fraud (misappropriation of assets, tax evasion, intentional mismarking of positions, bribery), External Fraud (theft of information, hacking damage, third-party theft and forgery), Employment Practices and Workplace Safety (discrimination, workers compensation, employee health and safety), Clients, Products, \& Business Practice (market manipulation, antitrust, improper trade, product defects, fiduciary breaches, account churning), Damage to Physical Assets (natural disasters, terrorism, vandalism), Business Disruption \& Systems Failures (utility disruptions, software failures, hardware failures), Execution, Delivery, \& Process Management (data entry errors, accounting errors, failed mandatory reporting, negligent loss of client assets).}.
There are therefore 56 combinations (8 business lines $\times$ 7 event types) to be implemented. 
The objective of our study is i) to accurately characterize these different losses, in order to ii) understand the main features of their dynamics and interactions, and if possible iii) minimize them.

In order to analyze the distribution of these losses, we shall model them via $N(=56)$ interacting variables, as proposed by   Kh\"{u}n and Neu \cite{kuhn} and elaborated by several authors in Refs.\ \cite{Leippold:2005,Clemente:2004,anand}.
This sets the afore-mentioned level of description at which techniques of spin glasses and complex systems can be applied.
An efficient performance of this model requires knowledge of the frequency and value of each loss in the $N$ different channels and a distribution of such events. Because Basel II is a rather recent treaty, real data are often insufficient. For example, some of the $8\times7$ channels are often completely empty, so that rare events cannot be included. Also, some business lines (e.g.\ Retail Banking) are often very populated, so that they tend to ``overwhelm" other lines when a statistical analysis is performed. Finally, the populations can be very different for different banks, and can show significant variations according to, e.g., bank size and geographical location. To focus, say, only on large banks, would not be meaningful, because one could aim at offering a service to \emph{many} small banks (say, minimize below 15\% the percentage of the total capital to be put aside), by taking into account their peculiarities and characteristics in a given local context. To add to these difficulties, banks are often reluctant to disclose the details of their losses, for obvious reasons. 
We shall therefore opt for a democratic approach, by setting the parameters of our model to reasonable values, valid for a rather wide range of banks, and checking that our results are stable against small variations of these parameters.

\section{The Model}
\label{sec:the_model}
The model we shall study consists of $N$ dichotomic interacting variables that evolve in time.
These variables schematize the $8 \times7$ interacting processes defined in the Introduction.
The model was introduced in Ref.\ \cite{kuhn}, elaborated in
\cite{Leippold:2005} and \cite{Clemente:2004}, where processes was proven to be dependent on each other, 
and further investigated by Kh\"{u}n and Anand in Ref.\ \cite{anand}, where correlations among processes were further explored. 
We shall adopt in the following a slightly different notation from the preceding articles. See Eqs.\  (\ref{eq:s_eta_mapping}) and (\ref{eq:evolution_equation_mapping}) in the following.

Let  $\vec{s}(t)$ be the $N$-dimensional vector of the dichotomic variables and $J_{ij}$ the coupling constant between $s_i$ and $s_j$. Time is discretized $t\in\mathbb{N}$ and $ s_i(t) \in \{ -1,+1 \}$, with   $i \in \{ 1, \dots, N \}$. 
The value $-1$ denotes a running process (no losses), while the value $+1$ a broken down process, that does not run anymore. 
In such a way, we are completely neglecting the amount of the loss and are only recording whether a loss occurs or not. In particular, since banks record only losses exceeding a given threshold \footnote{In Europe this threshold is 5000 EUR.}, we can interpret $s_i (t) = +1$ as the occurrence at time $t$ of a loss equal to or larger than this threshold in the $i$-th channel.

We assume that the time  evolution is governed by the equation 
\begin{equation}
\label{evolution_equation}
s_i(t+1) = \mbox{sign} \left[ \sum_{j=1}^N J_{ij}s_j(t) - \theta_i + \xi_i(t) \right] ,
\end{equation}
where $ \theta_i$ are characteristic constants, which we call \textit{supports}, and $\xi_i(t)$ are Gaussian noises
\begin{equation}
\label{gaussn}
\langle\xi_i(t) \rangle=0 \quad \mbox{and} \quad  \langle\xi_i(t)\xi_j(s) \rangle = \delta_{i,j} \delta_{t,s}\sigma^2_{\xi}.
\end{equation}
Here the brackets $\langle\cdot\rangle$ denote the average over the noise and $\delta_{i,j}$ is Kronecker's delta.

In order to understand the role of the parameters that appear in the evolution equation, let us decouple the processes, by setting $J_{ij}=0$  $\forall \; i,j$, so that Eq.\ (\ref{evolution_equation}) reads
\begin{equation}
 s_i(t+1) = \mbox{sign} \left[ -\theta_i + \xi_i(t) \right].
\end{equation}
The probability that $s_i(t+1)=+1$ is easily calculated 
\begin{equation}\label{support}
p_i=p_{ \{ s_i(t+1)=+1 \}} = \int_{\theta_i}^{\infty} \frac{\d\xi_i(t)}{\sqrt{2 \pi \sigma^2_{\xi}}}\; \e^{-\frac{\xi_i(t)^2}{2 \sigma^2_{\xi}}} ,
\end{equation}
that, after integration, yields 
\begin{equation}
p_i = \frac{1}{2} \left[ 1 - \mbox{erf}\Big(\theta_i / \sqrt{2\sigma^2_{\xi}}\Big) \right],
\label{eq:pitheta}
\end{equation}
where $\mbox{erf}(z)=2 \pi^{-1/2}\int_0^z \e^{-t^2/2} \; \d t$ is the error function.
(Notice that $s_i(t+1)$ is decoupled even from $s_i(t)$ in this simple example.)
We see that a given process $s_i$ runs after the time increment $t \to t+1$ if it is ``supported" by a suitable environment, that yields an average support $\theta_i (>0)$.
This elucidates the role of the supports for noninteracting variables: when nonvanishing couplings appear, the evolution is given by (\ref{evolution_equation}), so that a process keeps running over the time increment $t \to t+1$ if it is suitably supported by the argument of the sign function, which is a combination of its average support and the action of the other processes. Observe that the sign of the couplings $J_{ij}$ is not defined and that the average support $\theta_i$ is presumably defined by a process-dependent environment: a positive value $\theta_i>0$ denotes a certain ``care" of the bank that process $i$ keeps running.

The evolution equation can be given physical meaning by defining the following function of the configurations:
\begin{equation} \label{Hamiltonian}
H\left(t\right)= - \frac{1}{2}\sum_{i,j=1}^{N}{ J_{ij} s_{i}\left(t\right) s_{j}\left(t\right) + \sum_{i=1}^{N}{ \theta_{i} s_{i}\left(t\right) }}.
\end{equation}
If the couplings are randomly distributed, the above ``Hamiltonian" describes a spin glass in a site-dependent magnetic field.
If at each time step only one variable evolves (the so called \textit{asynchronous updating}), one can prove after some algebra that the function $\Delta H(t) = H(t+1) - H(t) $ is negative only if the following condition holds:
\begin{equation}
\label{simmetrico}
\mbox{sign} \left[ \sum_j J_{ij}s_j(t) - \theta_i \right] = \mbox{sign} \left[ \sum_{j \neq i} \frac{J_{ij}+J_{ji}}{2}s_j(t) - \theta_i \right].
\end{equation}
This means that, under the previous condition, $H(t)$ is a monotonously decreasing function of the configurations as the system evolves according to Eq.\ (\ref{evolution_equation}). This equation of motion can be independently derived from Eq.\ (\ref{Hamiltonian}):
\begin{equation}
\label{eq:conversion}
s_{k}\left(t+1\right)=\mbox{sign} \left[ - \frac{\partial H}{\partial s_{k}} \left(t\right) \right]. 
\end{equation}
If $H$ is viewed as a potential, $s_{k}$ aligns to the force derived from this potential. In the case of \textit{synchronous updating}, which is our case, one numerically observes that $\Delta H(t)$ is negative if the  coupling constants are averaged over. Observe that the condition (\ref{simmetrico}) is not entirely obvious in our context, where, say, a very lossy and highly populated channel might influence more a rarely populated channel than \emph{vice versa}.

In order to connect our results to those in \cite{anand}, notice that if 
we bijectively map the variables $s \in \{-1,+1\}$ into the variables $\eta \in \{0,+1\}$:
\begin{eqnarray}
s_i(t) &=& 2\eta_i(t) - 1 \in \{-1,+1\} , \nonumber \\
\eta_i(t) &=& \frac{1}{2} \left[ s_i(t) + 1 \right] \in \{ 0,1 \} ,
\label{eq:s_eta_mapping}
\end{eqnarray}
Eq.\ (\ref{evolution_equation}) reads
\begin{eqnarray}
 \eta_i(t+1)&=& \frac{1}{2}\mbox{sign}\left[ \sum_jJ_{ij}s_j(t) - \theta_i +\xi_i(t)\right] + \frac{1}{2}\nonumber\\ 
 &=& \Theta \left[ \sum_jJ_{ij}(2\eta_i(t)-1) -\theta_i +\xi_i(t)\right] \nonumber \\ 
  &=& \Theta \left[ \sum_j \tilde{J}_{ij}\eta_j(t)  -\tilde{\theta}_i + \xi_i(t) \right], 
\label{eq:evolution_equation_mapping}
\end{eqnarray}
where $\Theta$ is the Heaviside step function, and $\tilde{J}_{ij}=2 J_{ij}$ and $\tilde{\theta}_i=\theta_i + \sum_{j}J_{ij}$.

\section{Robustness of the model}
\label{sec:stability}

Before we start investigating the model introduced in the preceding section, it is useful to study its robustness when the parameters that characterize it are changed. This will also help us get a feeling of the way the dynamics is implemented and the different roles of the parameters. 

In order to study the evolution over $T$ time steps, let us define the vector of integers $\vec{z}(T)=(z_1(T),\dots, z_N(T))$, with
\begin{equation}
\label{zvec}
z_i(T)=\frac{1}{2}\sum_{t\leqslant T}(s_i(t)+1),
\end{equation}
that records the positive ($+1$) outcomes of the $\vec{s}=(s_1,\dots,s_N)$ variables during the whole evolution $T$. Note that the integers $z_i(T)$ are bounded by $0\leq z_i(T) \leq T$.
We ask the following question. 
Let $\vec{z}_*\in\mathbb{N}^N$ be a given vector (for instance the data set obtained by accumulating the losses of a given bank during a certain number of years).
How should we choose the parameters of the model dynamics in such a way that the evolution of the system yields $\vec{z}_*$ at time $T$? The simplest way to do this \cite{anand} is to assume that all variables are independent and define the probability that a variable takes the value $1$ according to
\begin{equation}
p_i
= \frac{ z_{*,i} }{ \sum_i z_{*,i} } . 
\end{equation}
In this way, we are assuming that these probabilities are constant in time. We then set in Eq.\ (\ref{evolution_equation}) all couplings $J_{ij}=0$ (independent variables), and from (\ref{eq:pitheta}) the supports
\begin{equation}
\theta_i = -\sqrt{2\sigma^2_{\xi_i}} \mbox{erfinv}\big(2p_i - 1 \big),
\end{equation}
where $\mbox{erfinv}$ is the inverse of the error function $\mbox{erf}$,
and let the system evolve for  $T= \sum_i z_{*,i}$ time steps. In this way we recover the given $\vec{z}_*$ on average. Indeed,
\begin{equation}
\langle z_i(T)\rangle= T p_i = z_{*,i}.
\end{equation}

We now introduce interactions among processes and consider the coupling constants as Gaussian random variables  with average and width
\begin{equation}
\label{quenchedn}
\overline{J_{ij}} = 0 \quad  \mbox{and}  \quad \overline{J_{ij}J_{kl}}  =\delta_{ik}\delta_{jl} \sigma^2_J,
\end{equation}
respectively, where the bar  denotes the average over the distribution of the $J_{ij}$'s. This is the familiar quenched picture: couplings change over a slow time scale and can be considered constant during the whole evolution, up to time $T$. The average must be performed at time $T$ over several evolutions, each with a different realization of the couplings.

\begin{table}[htdp]
\caption{Data set of accumulated losses $\vec{z}_*$ and corresponding 
supports $\vec{\theta}$ for $N=4$.  See upper panel of Fig.~\ref{fig:confront_an_qn_field0}.
They have mean $\mu_\theta= 0.7017$ and standard deviation $\sigma_\theta=0.8372$.}
\begin{center}
\begin{tabular}{c|rrrrr}
 \hline
 $\vec{z}_*$ & 6886 & 3047 & 56 & 11\\
\hline
$\vec{\theta}$ &  -0.2460 &  0.2555  & 1.2688 &  1.5287\\ 
  \hline
\end{tabular}
\end{center}
\label{tab2}
\end{table}
\begin{table}[htdp]
\caption{Data set of accumulated losses $\vec{z}_*$ and corresponding 
supports $\vec{\theta}$  for $N=10$.  See lower panel of Fig.~\ref{fig:confront_an_qn_field0}.
They have  mean $\mu_\theta= 0.9516$ and standard deviation $\sigma_\theta=0.5249$.}
\begin{center}
\begin{tabular}{c|rrrrr}
 \hline
 $\vec{z}_*$ & 4637&  2994 & 1648 & 239 & 209 \\
  & 90 & 30 & 90 & 60 & 3\\
\hline
$\vec{\theta}$ & 0.0456 &  0.2630  &  0.4874  & 0.9893   & 1.0174   \\
  & 1.1833 &  1.3744   & 1.1833 &   1.2566   &   1.7162\\
  \hline
\end{tabular}
\end{center}
\label{tab1}
\end{table}

\begin{figure}
 \begin{center}
  \includegraphics[width= 0.4 \textwidth]{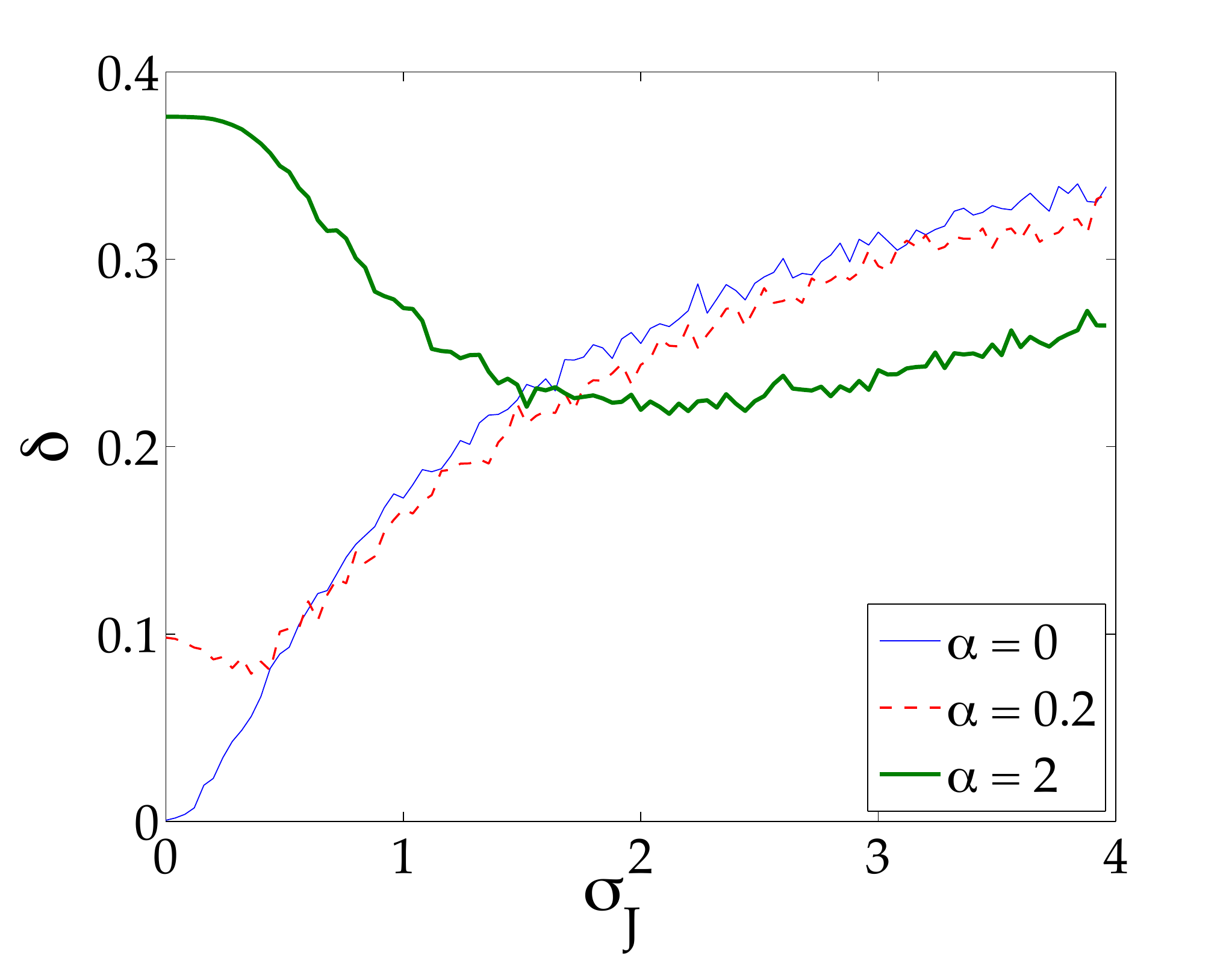} \\
  \includegraphics[width= 0.4 \textwidth]{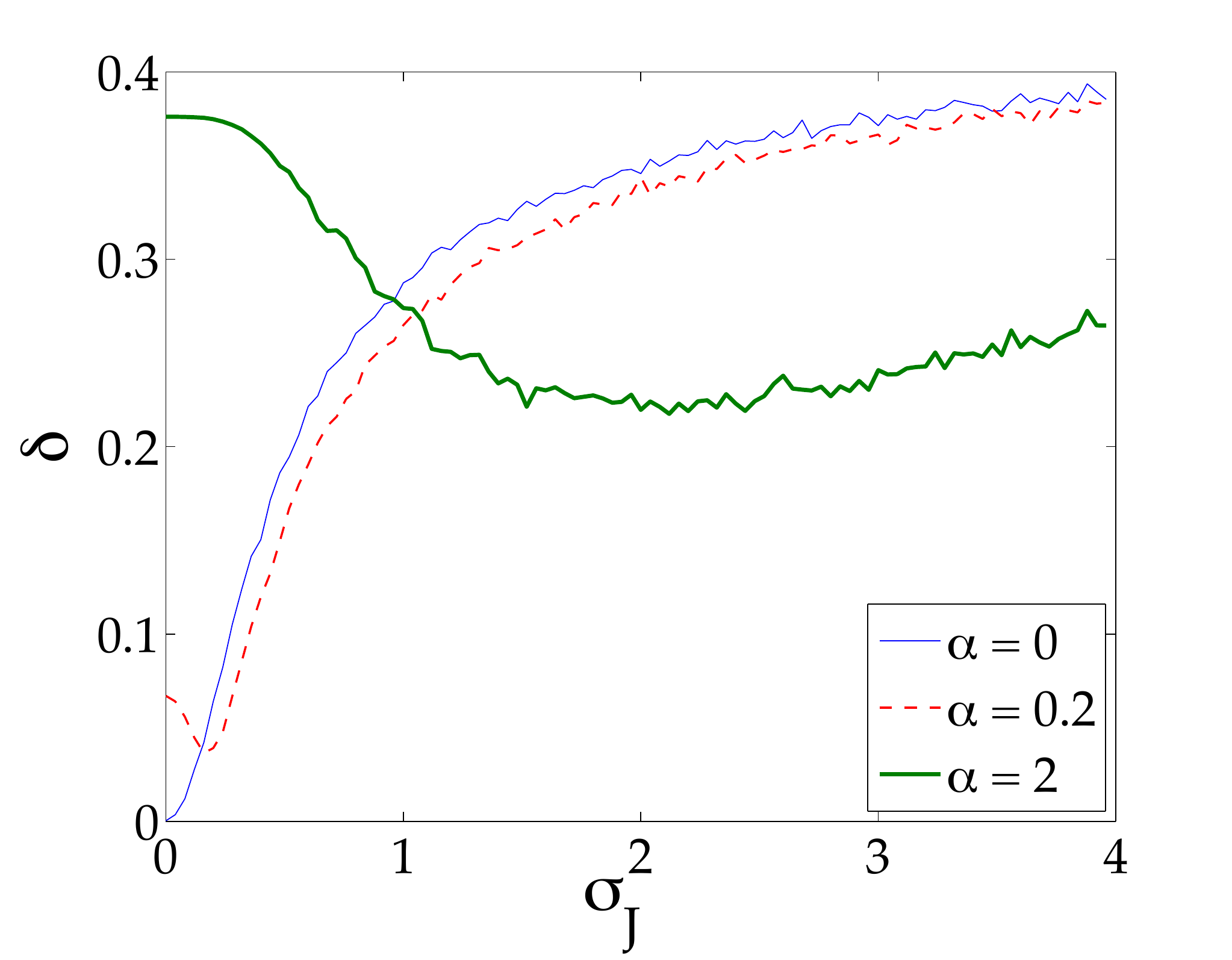}
     \caption{\label{fig:confront_an_qn_field0}(Color online) Distance $\delta$ in Eq.\ (\ref{eq:delta}) between the reference and final configurations vs standard deviation $\sigma_J$ of the distribution of the coupling $J_{ij}$. Upper panel $N=4$; lower panel $N=10$. 
Thin solid line: vanishing external field $\alpha$. The minimum is at $\sigma_J=0$, when all couplings vanish, and the final configuration is identical to the initial one; as $\sigma_J$ increases, the distance monotonically increases. 
Dashed line: $\alpha=0.2$; thick solid line $\alpha=2$. In both cases the distance is no longer zero at $\sigma_J=0$ because of the bias introduced by the field. As 
$\sigma_J$ increases, the spin interaction partially compensate for the external bias, so that
the distance reaches a minimum and then increases.}
\end{center}
\end{figure}

We start from the initial configuration $\vec{z}(0)=\vec{0}$ (all processes running) \footnote{We shall see in the following that the initial condition is basically irrelevant.} and numerically evaluate the distance
\begin{equation}
\label{eq:delta}
\delta = \frac{ \lVert  \overline{ \langle\vec{z}(T) \rangle} - \vec{z}_*\lVert  }{T \sqrt{N}} ,
\end{equation}
where the averages are defined in Eqs.\ (\ref{gaussn}) and (\ref{quenchedn}) and the norm is the Euclidean one, $\| \vec{a} \|=(\sum_i a_i^2)^{1/2}$. (Since $0\leq  \overline{ \langle z_i(T) \rangle}, z_{*,i}\leq T$,  the normalization is such that $0 \leq \delta \leq 1$.)
We look at two cases: $N=4$, with the supports shown in Table \ref{tab2}, and 
$N=10$, with the supports shown in Table \ref{tab1}.
The results are shown in Fig.\ \ref{fig:confront_an_qn_field0} (thin solid line $\alpha = 0$).
The evolution yields a final vector $\vec{z}(T)$ very close to $\vec{z}_*$ for $\sigma_J \simeq 0$, as expected. As 
$\sigma_J$ increases, the spins interact, causing the distance $\delta$ to increase monotonically. The result is qualitatively similar for both values of $N$. We observe that the evolution yields a vector that is rather close to $\vec{z}_*$ for sufficiently small couplings. 

In order to test the robustness of this result, let us add a constant external field $\vec{\alpha}=(\alpha, \dots, \alpha)$ to the supports
\begin{equation}
\vec{\theta} \rightarrow \vec{\theta} + \vec{\alpha}.
\end{equation}
We shall consider a small ($\alpha=0.2$) and a large ($\alpha=2$) perturbation of the supports $\vec{\theta}$. The presence of $\alpha$ introduces a bias (a systematic error) in the dynamics, and we ask whether it is possible to compensate for this error by suitably tuning the interactions. 
The dashed line ($\alpha = 0.2$) and the thick solid line ($\alpha = 2$) in Fig.\ \ref{fig:confront_an_qn_field0} answer this question: the distance is large in absence of spin interactions ($\sigma_J=0$). 
However, for larger $\sigma_J$, the distance reaches a minimum because some (random) couplings correct the evolution of the system in the desired direction. 

These results can be read in different ways. Interactions among processes can partially compensate for an erroneous choice of the supports. The very presence of a minimum (characterizing ``optimal" couplings) is an interesting feature of the model.
\emph{Vice versa}, strong interactions among processes can lead the dynamics astray, nullifying the stabilizing (beneficent) action of the supports.
The interplay among the different parameters will contribute to make the (highly nonlinear) dynamics a very interesting one.

\section{Analytical approximation}			
\label{sec:Analytical results}

We are now ready to tackle the problem in full generality.
Let us start by considering a one-step evolution, that can be analytically averaged. It is convenient to rewrite Eq.\ (\ref{evolution_equation}) in the form 
\begin{equation}
s_i(t+1) = 2 \Theta \left[ \sum_j J_{ij}s_j(t) - \theta_i + \xi_i(t) \right] - 1 .
\end{equation}
We now substitute the Heaviside function as the integral of its Laplace transform ($1/z$)
\begin{equation}
s_i(t+1) = \frac{1}{\pi \i }\int_{B}\frac{\d z}{z}\; \e^{z\left[\sum_jJ_{ij}s_j(t) - \theta_i + \xi_i(t)\right]} -1
\end{equation}
along a vertical line $B$ just at the right of the imaginary axis (Bromwich's path).

We take the average over the noise and the couplings
\begin{widetext}
\begin{eqnarray}
\overline{\langle s_i(t+1) \rangle }&=&  \int \prod_j \left[\frac{\d J_{ij}}{\sqrt{2\pi \sigma^2_J}}\, \e^{-\frac{J_{ij}^2}{2\sigma^2_J}} \right]  \int\frac{\d\xi_i}{\sqrt{2\pi\sigma^2_{\xi}}}\e^{-\frac{\xi_i^2}{2\sigma^2_{\xi}}}  \frac{1}{\pi \i}\int_{B}\frac{\d z}{z}\, \e^{z\left[\sum_jJ_{ij} s_j(t) + \xi_i-\theta_i\right]}-1 .
\end{eqnarray}
\end{widetext}
Gaussian integrations yield
\begin{equation}\label{bromwich_integral}
\overline{\langle s_i(t+1) \rangle }=\frac{1}{\pi \mbox{i}}\int_{B}\frac{\d z}{z}\;\e^{\frac{z}{2}(z\sigma^2_{\xi}-2\theta_i)+\frac{1}{2}z^2\sigma^2_J \sum_j s_j^2(t)}-1.
\end{equation}
On the Bromwich path, the integration variable takes the form $z=\gamma+\mbox{i} y$ with $\gamma > 0 $; the explicit form of the previous integral is then
\begin{eqnarray}
&&\overline{\langle s_i(t+1) \rangle }= \frac{\e^{ \frac{1}{2}\gamma^2 k -\gamma\theta_i }}{\pi} \int^{+\infty}_{-\infty}\frac{\d y}{\gamma + \mbox{i}y}\; \e^{-\frac{1}{2}y^2\sigma^2_{\xi}} \nonumber \\  &&\times\left[ \cos\left[ \gamma y k -y\theta_i \right] + \i\sin\left[ \gamma y k-y\theta_i \right] \right] -1,
\end{eqnarray}
where 
\begin{equation}
k = \sigma^2_{\xi} + \sigma^2_{J} \sum_j s^2_j(t) = \sigma^2_{\xi} + N \sigma^2_{J}.
\end{equation}
In the limit $\gamma \rightarrow 0$, the first contribution yields
\begin{eqnarray}
&&\frac{1}{\pi}\int_{-\infty}^{\infty}\d y\;\frac{-\i}{y-\i\gamma}\,\e^{-\frac{1}{2}y^2\sigma^2_{\xi}}\cos\left[ \gamma y k -y\theta_i \right]  \nonumber\\  
&&=P\frac{-\i}{\pi}\int_{-\infty}^{\infty}\frac{\d y}{y}\;\e^{-\frac{1}{2}y^2\sigma^2_{\xi}}\cos\left[ y\theta_i \right]+ 1,
\end{eqnarray}
in which we have used the well-known 
formula for distributions
\begin{equation}
 \frac{1}{x-(x_0 \pm \mbox{i}\epsilon)} =P\,\frac{1}{x-x_0}\pm\i\pi\delta (x-x_0) ,
\end{equation}
with $P$ denoting the Cauchy principal value.
Due to symmetry considerations, the previous integral yields a contribution $1/2$ to the total integral. The remaining part does not have a pole, so we can set $\gamma =0$ in the integral, obtaining 
\begin{equation}
\frac{1}{\pi}\int_{-\infty}^{\infty}\d y\; \frac{\sin\left( -y\theta_i \right)}{y}\, \e^{-\frac{1}{2}y^2\sigma^2_{\xi}} = -\mbox{erf}\left( \frac{\theta_i}{\sqrt{2k}}\right).
\end{equation}
Summing up,
\begin{eqnarray}
\label{annealed_evolution}
\overline{\langle s_i(t+1) \rangle} 
&=& \mbox{erf} \left[\frac{-\theta_i}{\sqrt{2(\sigma^2_{\xi}+N \sigma^2_J )}} \right] .
\end{eqnarray}
It is clear that by iterating this expression over the whole evolution $[0,T]$ one makes a mistake, because the average over the couplings is taken at every time step. 

Another alternative would be to compute the noise-averaged single-step evolution
\begin{eqnarray}
\langle s_i(t+1) \rangle &=& \int \frac{d\xi_i}{\sqrt{2\pi\sigma^2_{\xi}}}\;\e^{-\frac{\xi^2_i}{2\sigma^2_{\xi}}} \nonumber \\ & &\times \mbox{sign} \left[ \sum_{j=1}^N J_{ij}s_j(t) - \theta_i + \xi_i \right] \nonumber \\
 &=&\mbox{erf}\left[ \frac{ \sum_jJ_{ij}s_j(t) -\theta_i }{\sqrt{2\sigma^2_{\xi}}} \right]
 \label{quenched_evolution}
 \end{eqnarray}
and average over the realizations of the couplings at the final step. This also entails an error, because we are forcing a Markovian evolution.
 
\begin{figure}
\centering
\includegraphics[width=0.5 \textwidth]{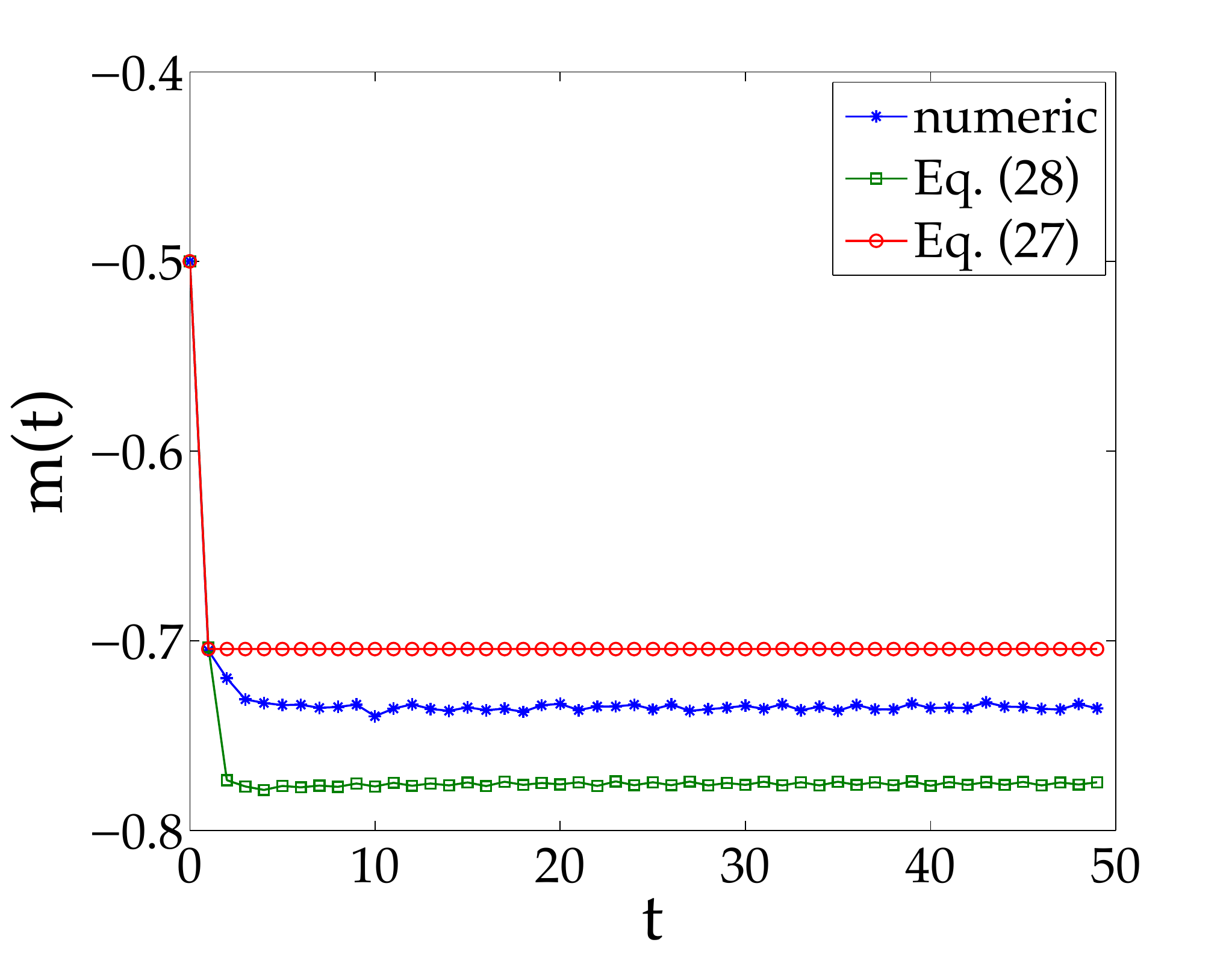}
\caption{\label{fig:confront}(Color online) Evolution of the magnetization. Comparison between the numerical evolution \eqref{evolution_equation}, and the approximations \eqref{annealed_evolution} and \eqref{quenched_evolution}. Notice that $m(0) = 0.5$.}
\end{figure}

\begin{figure}
\centering
\includegraphics[width=0.5 \textwidth]{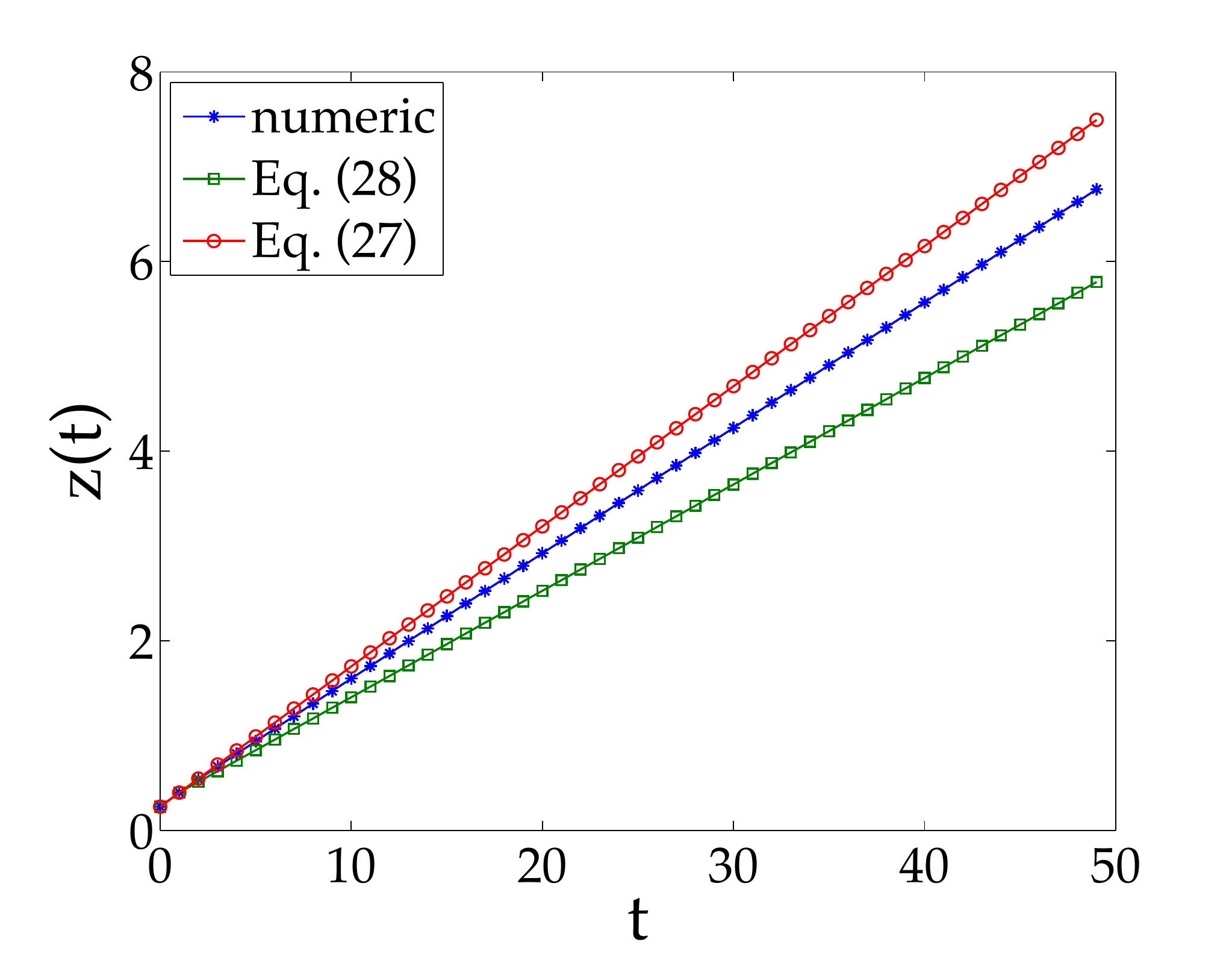}
\caption{\label{fig:z_confront}(Color online) Evolution of the quantity $z(t) = \sum_i z_i(t)$, see Eq.\ \eqref{zvec}. Comparison between the numerical evolution \eqref{evolution_equation}, and the approximations \eqref{annealed_evolution} and \eqref{quenched_evolution}.}
\end{figure}

We shall now compare the expressions \eqref{annealed_evolution} and \eqref{quenched_evolution} to the numerical evolution \eqref{evolution_equation}. In particular we will look at the quantity
\begin{equation}
\label{mt}
 m(t) = \frac{1}{N}\sum_{i=1}^N \overline{\langle s_i(t) \rangle },
\end{equation}
which is simply the sum of the mean components of the spin vector $\vec{s}$ and has the (appealing) physical meaning of average magnetization. This quantity will be thoroughly analyzed in the remaining part of this article.

The three evolutions in Fig.\ \ref{fig:confront} are similar, in that the magnetization tends in all cases to an asymptotic value
\begin{equation}
\label{mas}
m = \lim_{t \rightarrow \infty}m(t) .
\end{equation}
These asymptotic values coincide within 10\%.
Notice, however, that this error (linearly) accumulates in Eq.\ (\ref{zvec}), as shown in Fig.\ \ref{fig:z_confront}. 
Moreover, both approximations \eqref{annealed_evolution} and \eqref{quenched_evolution} are unable to describe the features of the short-time behavior. For this reason, we shall numerically evaluate the evolution of the magnetization \eqref{mt}, by using \eqref{evolution_equation}.

\section{Asymptotic magnetization}
\label{sec:Asymptotic_magnetization}

\begin{figure}
 \begin{center}
 \includegraphics[width= \columnwidth]{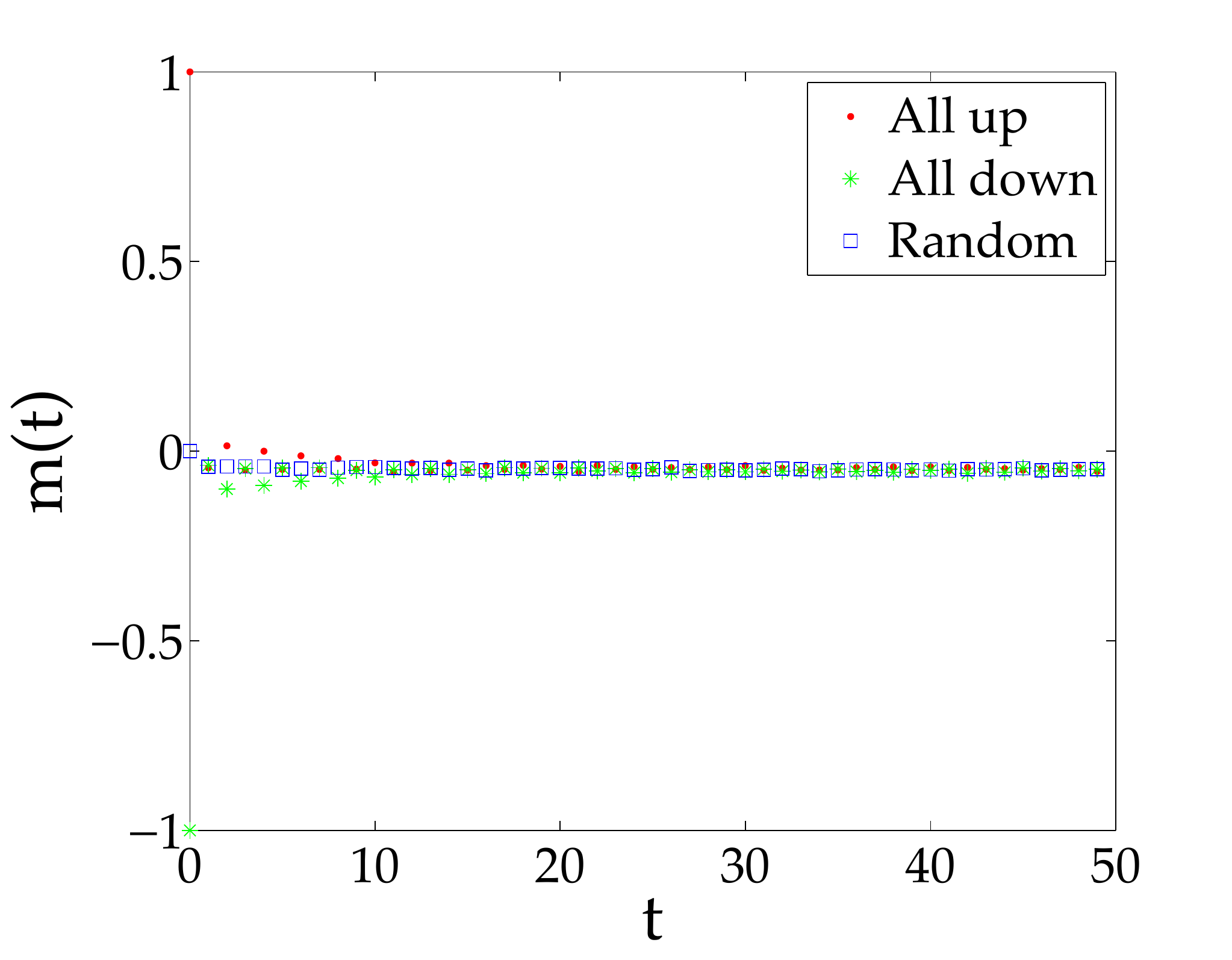}
 \caption{\label{fig:confront_over_initial_condition} (Color online) Time evolution of the magnetization of the system for three different initial configurations: all spins up (dots), all spins down (stars) and randomly oriented spins (squares). The couplings are quenched random variables, chosen according to \eqref{quenchedn}, while the supports $\theta_i \in \left[ -1 , 1 \right]$ with $\mu_\theta=0.1746$ and $\sigma_\theta=0.5870$.}
 \end{center}
 \end{figure}

The magnetization (\ref{mt}) and its asymptotic value (\ref{mas}) clearly depend on the details of the model. However, the initial spin configuration has practically no influence, as shown in Fig.\ \ref{fig:confront_over_initial_condition}, that displays the evolution of $m(t)$ starting from three different initial configurations: one in which all processes are initially running, $s_i(0) = -1 \; \forall i$, one in which they are all broken down, $s_i(0)=1 \; \forall i$, and one in which they are randomly distributed.
We took  $N = 10$, $\sigma_{\xi} = \sigma_J = 1$ and the supports (randomly) distributed in $[-1,1]$, namely $\vec{\theta}=\;$(0.3115,   -0.9286,    0.6983,    0.8680,    0.3575,    0.5155,    0.4863,   -0.2155,    0.3110,   -0.6576), with mean 
\begin{equation}
\label{eq:meantheta}
\mu_\theta=\frac{1}{N} \sum_{i=1}^N \theta_i
\end{equation}
equal to $0.1746$ and standard deviation 
$\sigma_\theta=0.5870$. Clearly, the system quickly forgets its starting point, and the magnetization always converges to a common state that does not depends on the initial configuration. For this reason  in the following analysis we will always consider an initial configuration in which all spins are down.
 
By contrast, the magnetization strongly depends on the supports and on the standard deviations of the distributions of the couplings and the noise. 
When $\sigma_{\xi}$ and $\sigma_J$ are very large, the asymptotic magnetization always tends to $0$, independently of the supports. This can be seen in all the figures that follow and is in agreement with intuition: when noise is large and the distribution of couplings has a large standard deviation, the role of supports becomes negligible in Eq.\ (\ref{evolution_equation}), and 
the spins take the values $\pm1$ with equal probability.

\begin{figure}
 \begin{center}
 \includegraphics[width= \columnwidth] {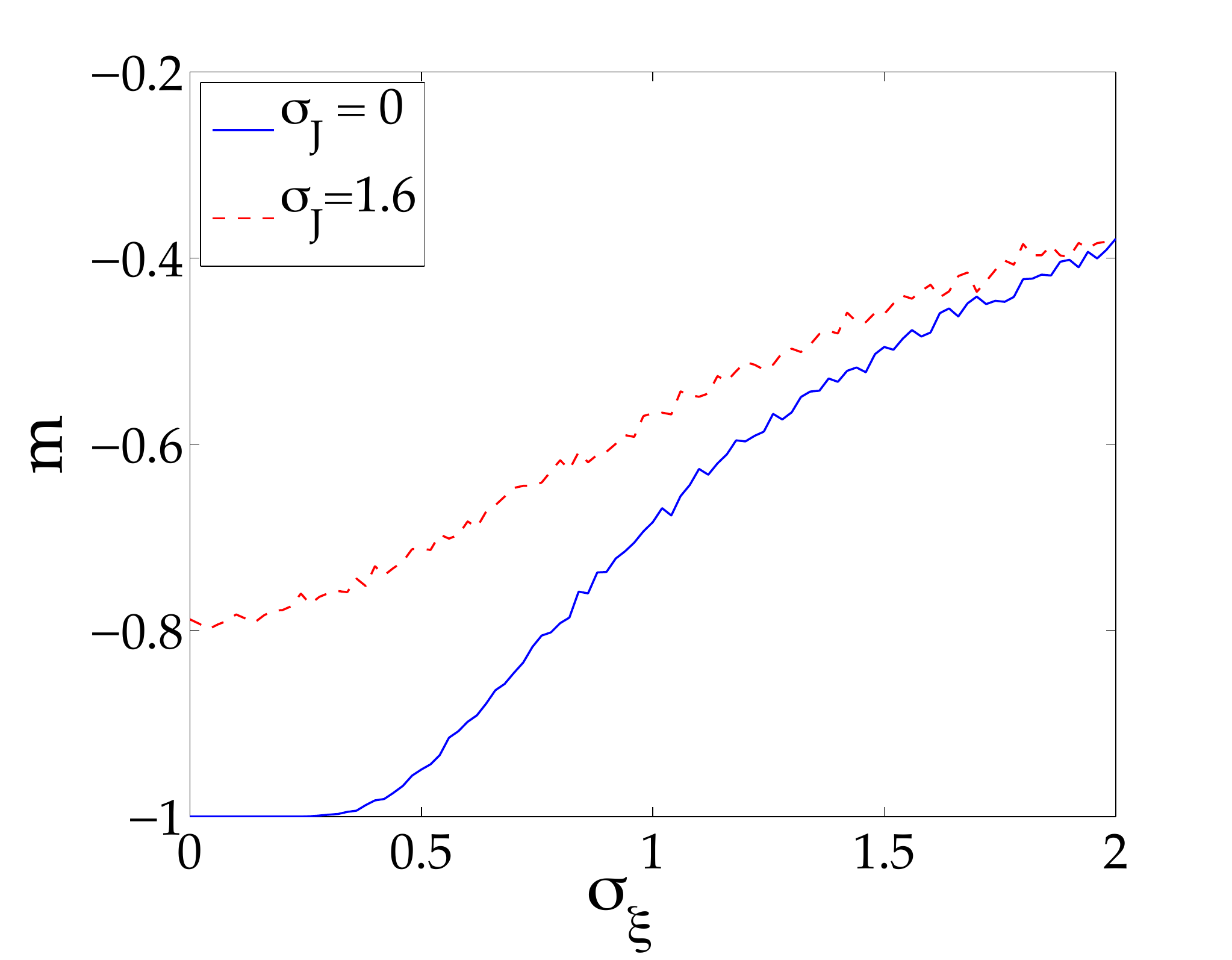}
 \caption{(Color online) Asymptotic magnetization vs $\sigma_{\xi}$. The couplings are quenched random variables, chosen according to \eqref{quenchedn}, while the supports $\theta_i \in \left[ 4 , 6 \right]$ with $\mu_\theta=5.1381$ and $\sigma_\theta=0.7841$. Solid line: $\sigma_J=0$ (no couplings); dashed line: $\sigma_J=1.6$.} 
 \label{fig:av_theta_noiseless_and_independent}
 \end{center}
\end{figure}

Let us first consider the case of $N=5$ spins and large positive supports in the interval $[ 4 , 6 ]$, namely
$\vec{\theta} = (4.1951, 4.5570, 5.0938, 5.9150, 5.9298)$ ($\mu_\theta=5.1381$ and $\sigma_\theta=0.7841$). 
We first consider the 
non-interacting case ($\sigma_J = 0$).
As $\sigma_\xi$ increases, the spins can make transitions to the $+1$ state
and the asymptotic magnetization increases.
Since all supports are positive, the standard deviation of the noise $\sigma_{\xi}$ must exceed a certain threshold before transitions can take place. See solid line in Fig.\ \ref{fig:av_theta_noiseless_and_independent}.
On the other hand, for strongly interacting spins  $\sigma_J=1.6$, the effect of noise is essentially negligible, see dashed line in Fig.\ \ref{fig:av_theta_noiseless_and_independent}. Clearly, both lines tend to 0 as  $\sigma_{\xi} \to \infty$ (not shown).

\begin{figure}
 \begin{center}
 \includegraphics[width= \columnwidth]{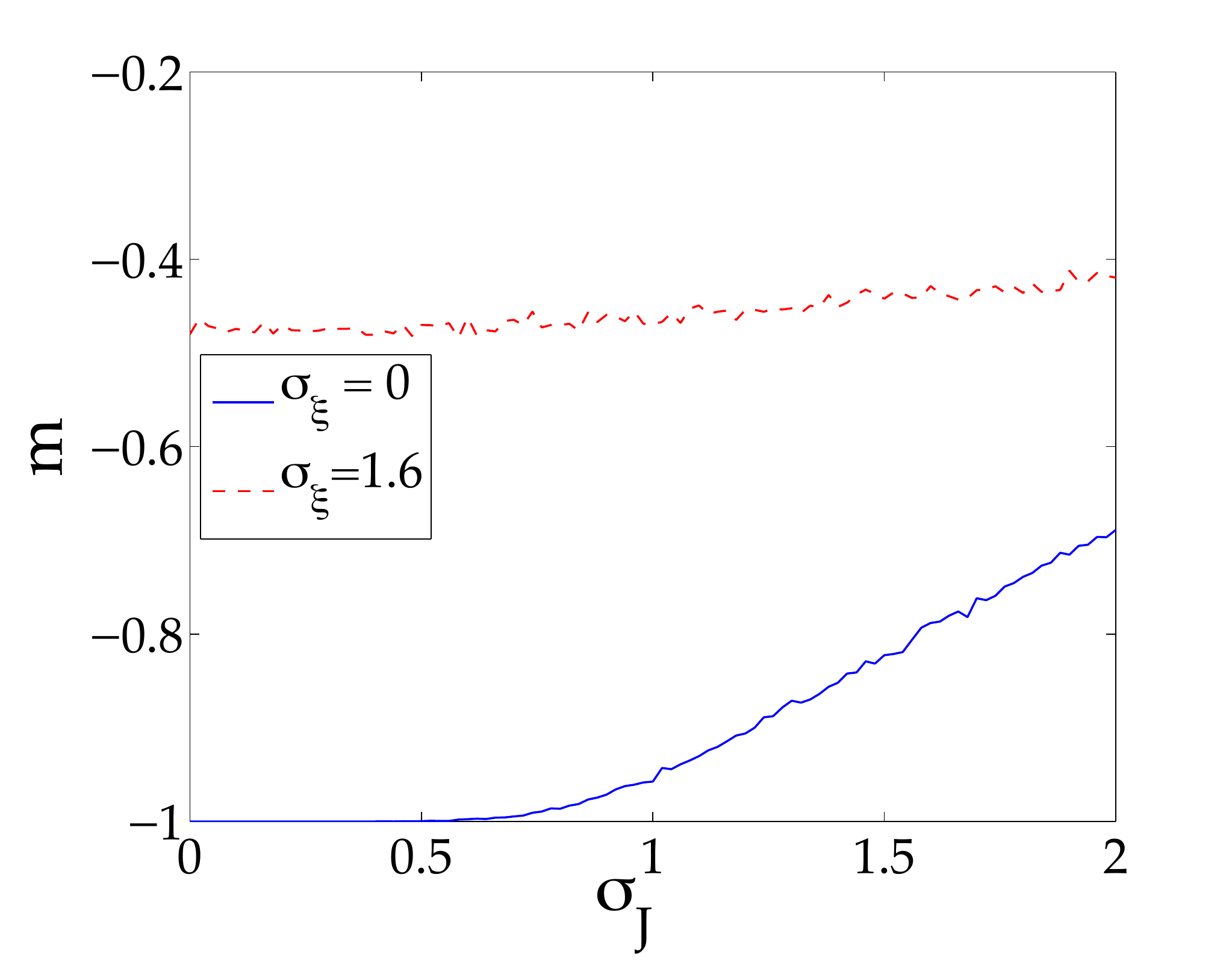}
 \caption{ \label{fig:asymptotic_magnetization_noisefull_dependent} (Color online) Asymptotic magnetization vs $\sigma_J$. The couplings are random quenched variables chosen according to \eqref{quenchedn} while the supports $\theta_i \in \left[ 4 , 6 \right]$ with $\mu_\theta=5.1381$ and $\sigma_\theta=0.7841$. Solid line: $\sigma_\xi=0$ (noiseless case); dashed line  $\sigma_\xi=1.6$.}
 \end{center}
\end{figure}

We now perform the same analysis by exchanging the role of the two standard deviations. 
Figure \ref{fig:asymptotic_magnetization_noisefull_dependent} displays the behavior of the same system with large positive supports in $[ 4 , 6]$, 
as $\sigma_J$ varies. 
The solid line is the evolution in the noiseless case, while the dotted line is the evolution for $\sigma_{\xi}=1.6$. The behavior is qualitatively similar to that in Fig.\ \ref{fig:av_theta_noiseless_and_independent}, but quantitatively different. We shall come back to this point in the next section.

\begin{figure}
 \begin{center}
 \includegraphics[width=\columnwidth]{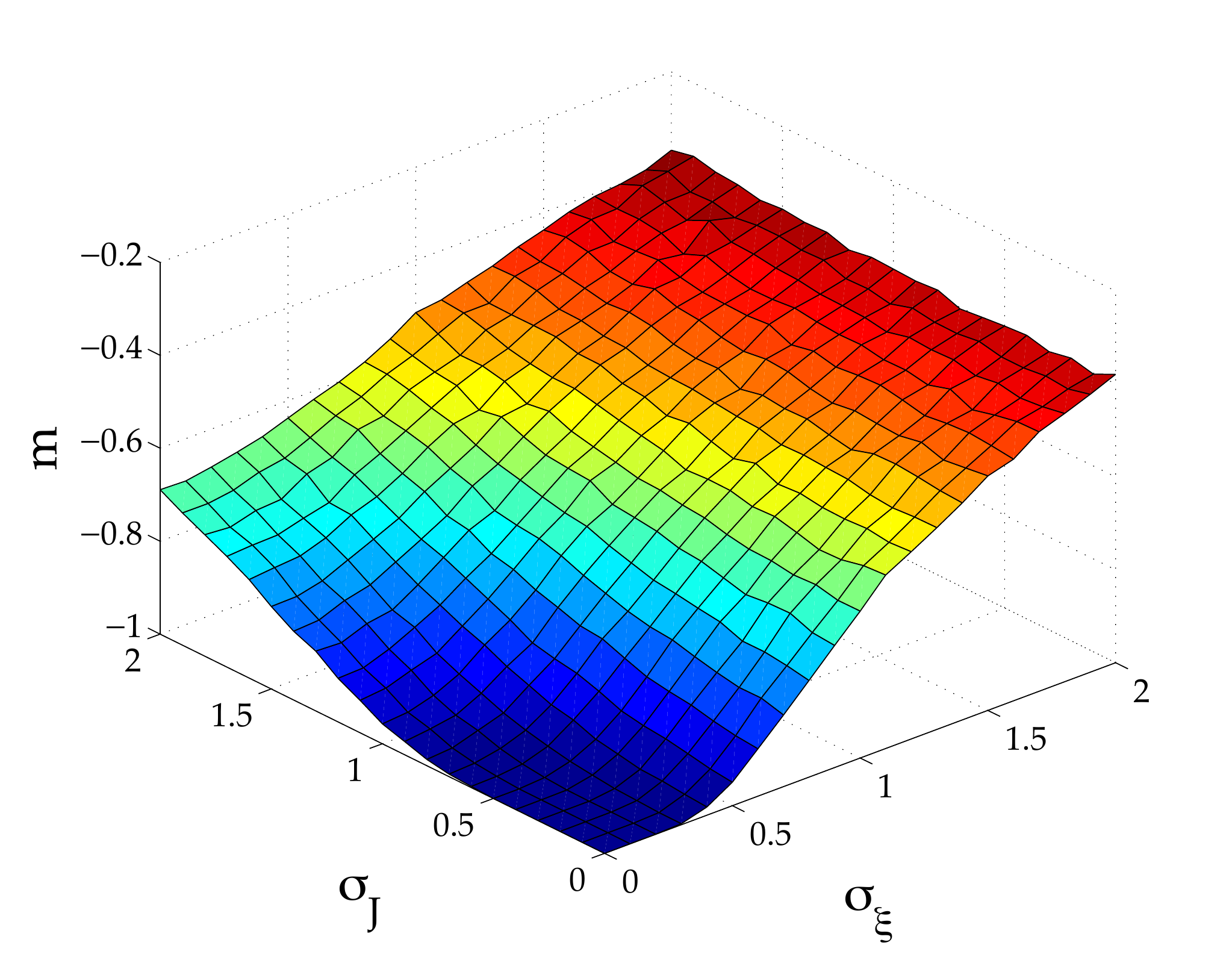}
 \caption{(Color online) Asymptotic magnetization vs $\sigma_J$ and $\sigma_{\xi}$. The couplings are quenched random variables, chosen according to \eqref{quenchedn}, while the supports $\theta_i \in \left[ 4 , 6 \right]$ with $\mu_\theta=5.1381$ and $\sigma_\theta=0.7841$. The magnetization tends to zero as $\sigma_{\xi}$ and/or $\sigma_J$ become large.}
 \label{fig:3D_qn_scsi_sgei_theta5}
 \end{center}
 \end{figure}

These conclusions are summarized in Fig.\  \ref{fig:3D_qn_scsi_sgei_theta5}.
Notice the asymmetry $\sigma_{\xi} \leftrightarrow \sigma_J$. Observe that 
the magnetization tends to vanish as $\sigma_{\xi}$ and/or $\sigma_J$ become large.
Notice also that due to the symmetry of the problem, the situation with large negative supports would be identical (modulo a reflection $m \leftrightarrow -m$).

\begin{figure}
 \begin{center}
 \includegraphics[width=\columnwidth]{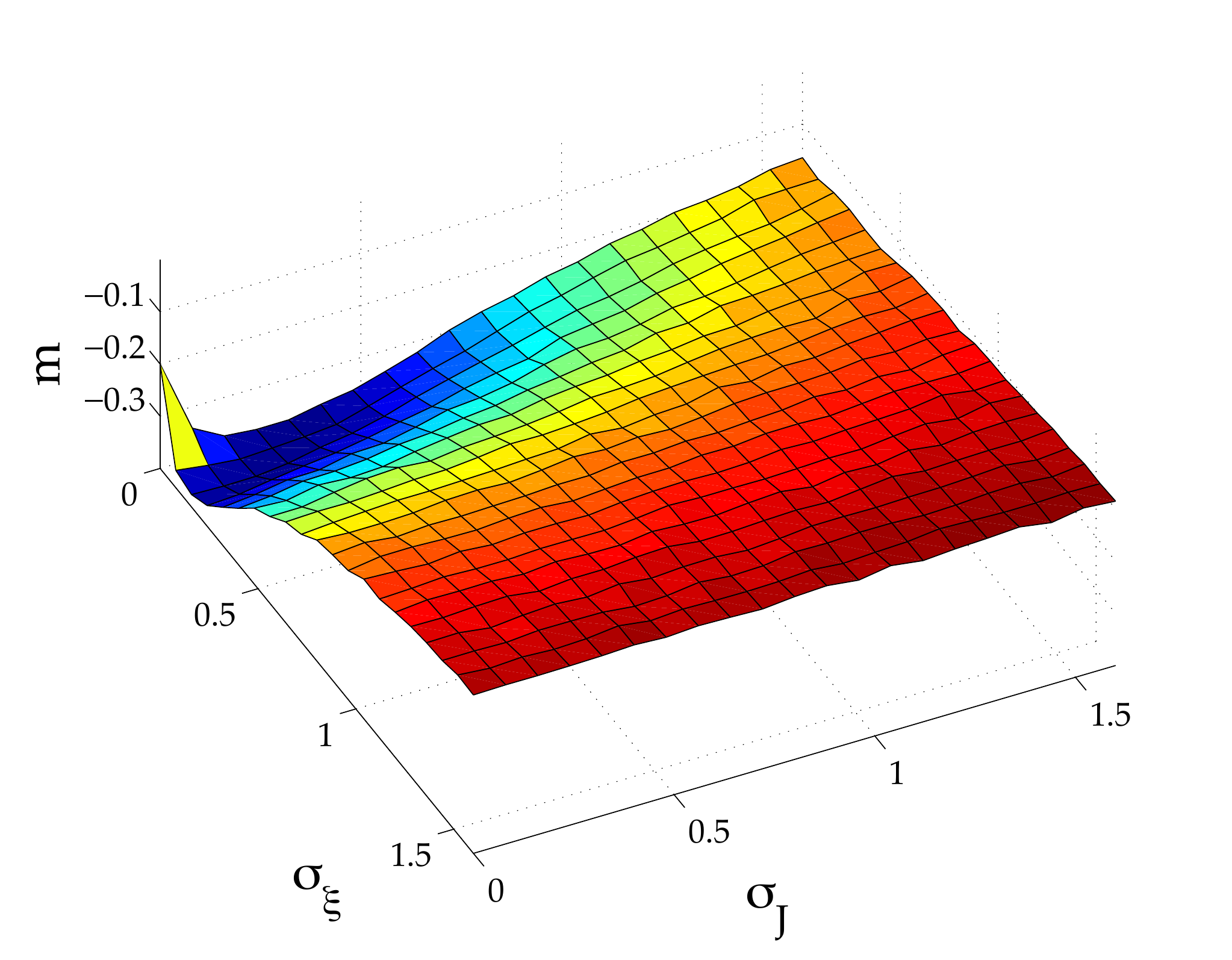}
 \caption{(Color online) Asymptotic magnetization vs $\sigma_J$ and $\sigma_{\xi}$. The couplings are quenched random variables, chosen according to \eqref{quenchedn}, while the supports $\theta_i \in \left[ -1 , 1 \right]$ with $\mu_\theta=0.3484$ and $\sigma_\theta=0.6974$. The magnetization tends to zero as $\sigma_{\xi}$ and/or $\sigma_J$ become large.}
 \label{fig:3D_qn_scsi_sgei_theta0}
 \end{center}
 \end{figure}

The scenario changes when the supports are randomly chosen in the interval $\left[ -1,1 \right]$, e.g.\
$\vec{\theta} = (-0.6848,  0.9412,    0.9143,   -0.0292,   0.6006)$
($\mu_\theta=0.3484$ and  $\sigma_\theta=0.6974$). Figure \ref{fig:3D_qn_scsi_sgei_theta0} shows the asymptotic magnetization
when $\sigma_J$ and $\sigma_{\xi}$ vary. As one can see, the magnetization (quickly) tends to a minimum and then slowly tends to its asymptotic value $0$.

\begin{figure}
 \begin{center}
 \includegraphics[width=\columnwidth]{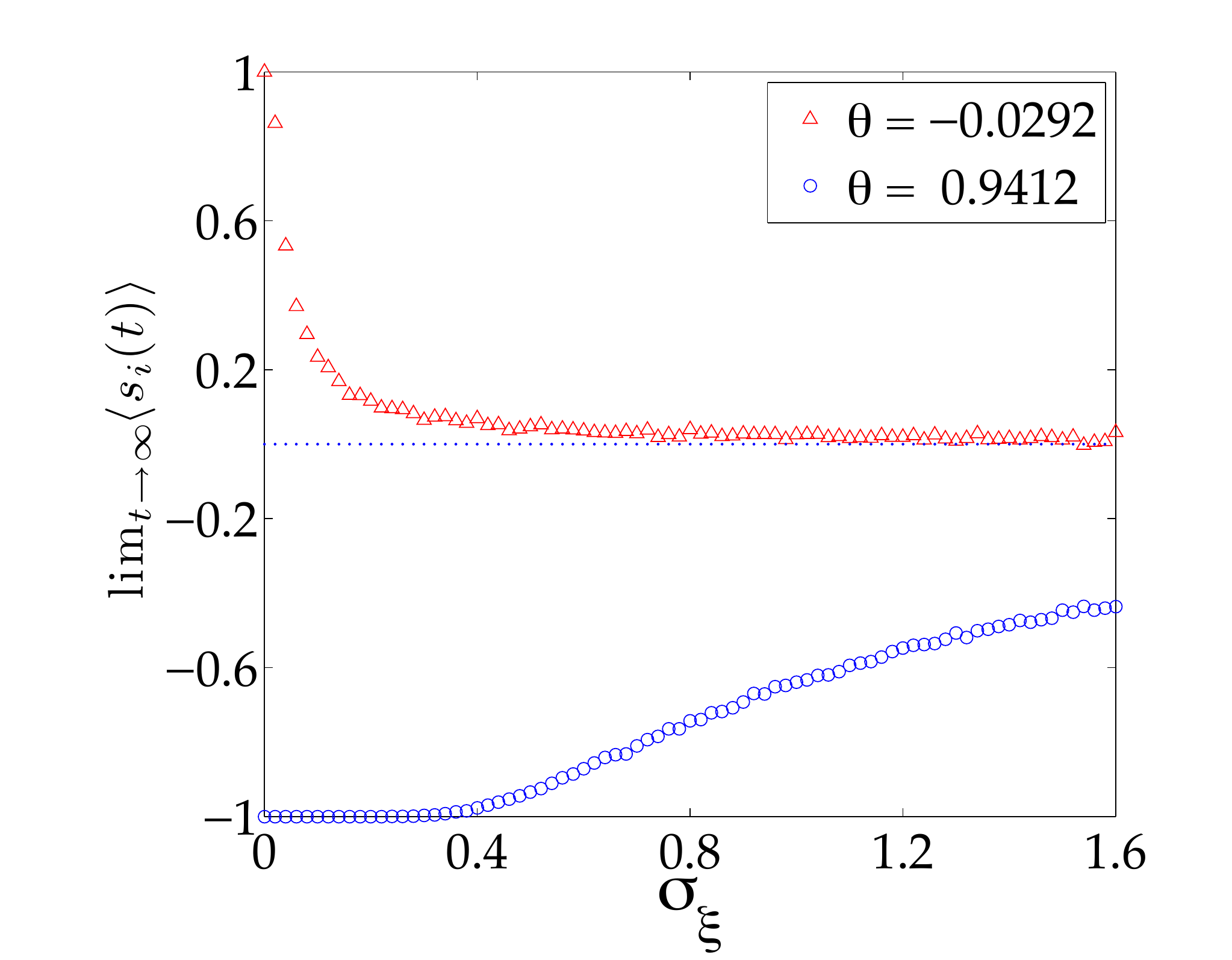}
 \caption{\label{fig:confront_high_low_theta}(Color online) 
 Asymptotic average value of two particular spins vs $\sigma_{\xi}$. The two spins have very different supports $\theta$. The couplings are quenched random variables, chosen according to \eqref{quenchedn}. When $|\theta|$ is small the convergence is very fast, not so when it is larger. By summing curves like those in this figure one obtains the minimum that appears in Fig.~\ref{fig:3D_qn_scsi_sgei_theta0}.}
  \end{center}
 \end{figure}

The plot in Fig.\ \ref{fig:confront_high_low_theta} illustrates this aspect in more detail. Let us consider the case with no couplings, $\sigma_J=0$, and look at  the asymptotic mean value of two \emph{particular} spins: one with a large (positive) support $\theta=0.9412$ and one with a small (negative) support $\theta=-0.0292$. When $|\theta|$ is small, a small noise $\sigma_\xi$ is sufficient to drive the mean value of the spin to zero. On the other hand, a spin with a large support is rather insensitive to small values of noise $\sigma_\xi$, which should therefore be sufficiently high in order to drive it towards 0.
By summing these different contributions to the asymptotic magnetization one obtains the minimum in Fig.\ \ref{fig:3D_qn_scsi_sgei_theta0}. More in general, the behavior of the magnetization for small values of $\sigma_\xi$ and/or $\sigma_J$ strongly depends on the value of $\theta_i$'s, while (obviously) as the standard deviations increase, all variables tend to the same asymptotic value, and this yields the flat part of the graph. This qualitatively explains all the features of Fig.\  \ref{fig:3D_qn_scsi_sgei_theta0}.
We remind here that $s_i=+1$ represents a loss in channel $i$, while $s_i=-1$ means that no loss occurs.


\section{Asymptotic Behaviour}
 \label{sec:asymptotic_behaviour}

\begin{figure}
 	\centering
 	\includegraphics[width=\columnwidth]{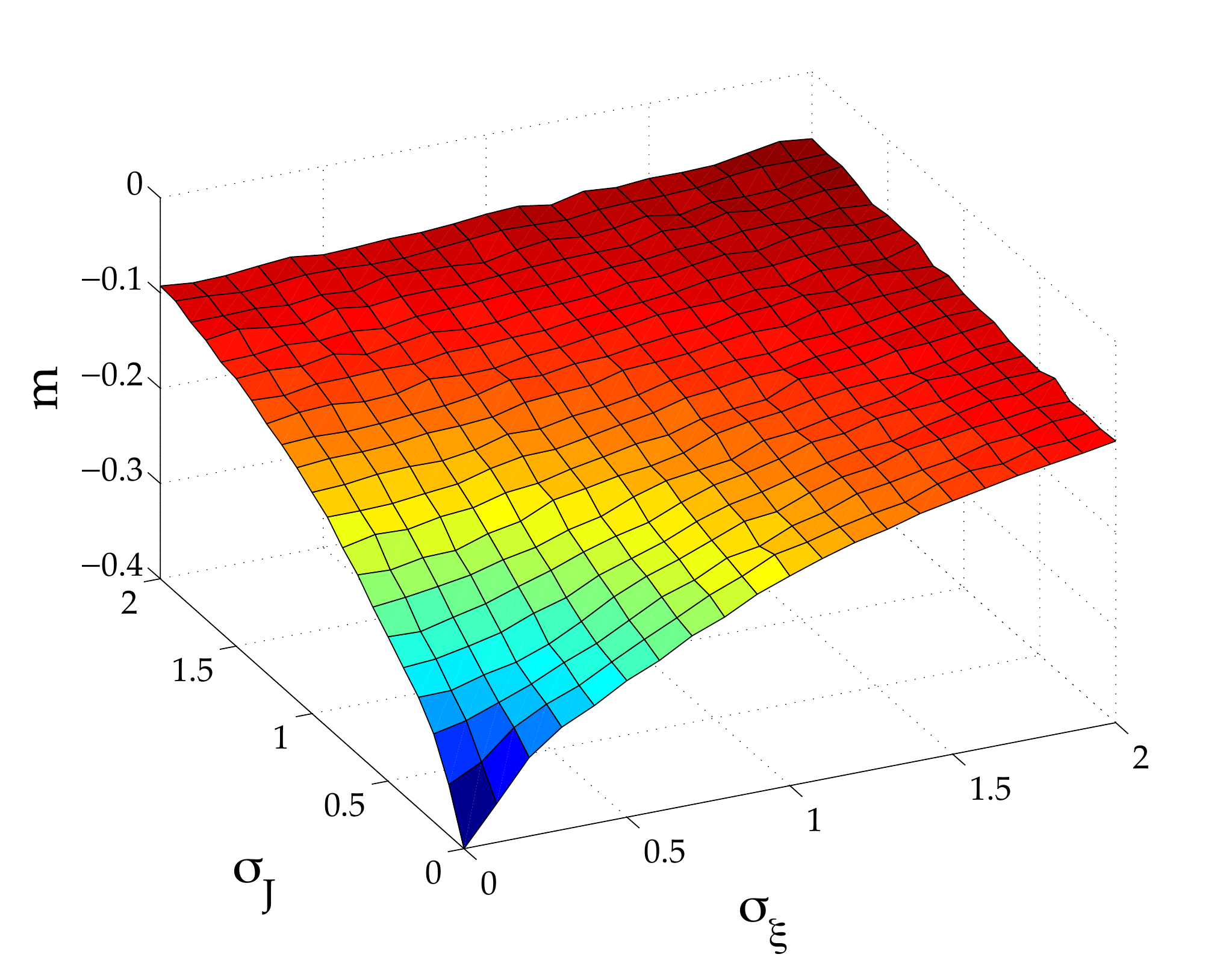}
 	\caption{\label{fig:magnetizzazione_asintotica_conflittuale}(Color online) Asymptotic magnetization vs $\sigma_J$ and $\sigma_{\xi}$. The couplings are chosen according to \eqref{eq:funzione_distribuzione}, while the supports $\theta_i \in \left[1 , 2 \right]$ with $\mu_\theta=0.2477$ and $\sigma_\theta=0.6917$. The magnetization tends to zero as $\sigma_{\xi}$ and/or $\sigma_J$ become large.}
 \end{figure}

In this section, at variance with the previous ones, we will consider couplings with a nonvanishing mean. However, we will appropriately scale the mean and standard deviation of the couplings in term of the lattice size $N$ and the mean support $\mu_\theta$ in Eq.\ (\ref{eq:meantheta}) in order to disclose an interesting behavior.
We take
 \begin{eqnarray} \label{eq:funzione_distribuzionea}
	& & P(\xi_i(t)) =  \mathcal{N} \left( 0,\sigma_{\xi} \right), \nonumber\\
 	& & P(J_{ij}) = \mathcal{N} \left(- \frac{\mu_\theta}{N},\frac{\sigma_{J}}{\sqrt{N}} \right),
\label{eq:funzione_distribuzione}
\end{eqnarray}
where $\mathcal{N}(\mu,\sigma)$ denotes the Gaussian probability density with mean $\mu$ and standard deviation $\sigma$. 
The behavior of the asymptotic magnetization in the scaled variables is shown in Fig.\  \ref{fig:magnetizzazione_asintotica_conflittuale} for $N=10$. We take randomly distributed $\theta_i \in [1,2]$, namely $\vec{\theta}=\;$(0.6294,    0.8116,   -0.7460,    0.8268,    0.2647,   -0.8049,   -0.4430,    0.0938,    0.9150,    0.9298)  ($\mu_\theta=0.2477$ and $\sigma_\theta=0.6917$).
Let us explicitly observe that such value of $\mu_\theta$ introduces in (\ref{eq:funzione_distribuzione}) an antiferromagnetic bias in the couplings. 
Remarkably, the graph displays a symmetry  $\sigma_J \leftrightarrow\sigma_{\xi}$ as a consequence of the performed scaling.

\begin{figure}
 	\centering
 	\includegraphics[width=\columnwidth]{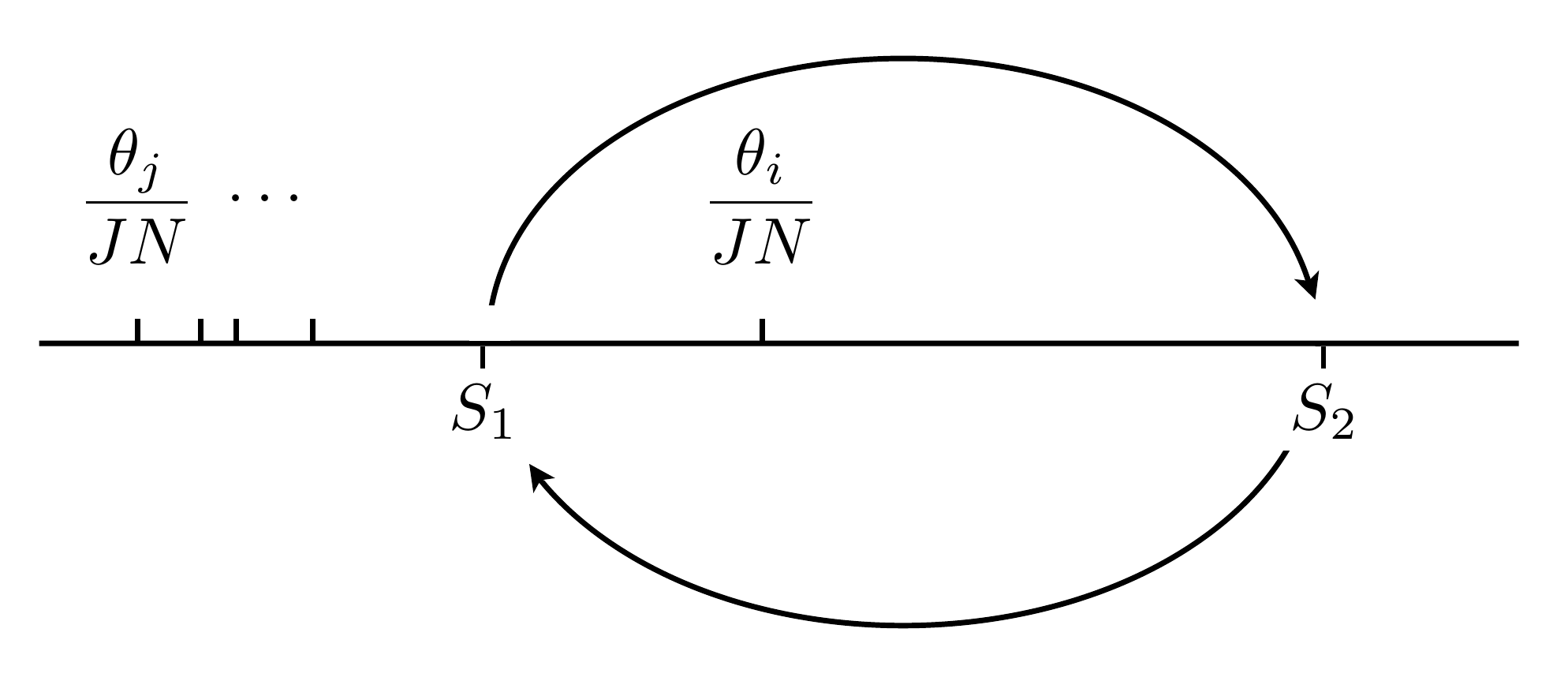}
 	\caption{\label{fig:oscillazione}(Color online) Case in which all the supports satisfy the condition $\theta_j/JN<S_1$, except the $i$-th one, that satisfies the condition $S_1<\theta_i/JN<S_2$. The $i$-th spin flips, and the magnetization indefinitely oscillates between the values $-1$ and $-1 +2/N$.}	
\end{figure}

We now scrutinize an interesting feature of the evolution equation (\ref{evolution_equation}).
In the noiseless case this yields
 \begin{equation}
 	s_i(t+1) = 
 	\begin{cases}
 		+1 & \text{if} \quad \sum_{j=1}^N J_{ij}s_j(t) -\theta_i >0, \\
 		-1 & \text{if} \quad  \sum_{j=1}^N J_{ij}s_j(t) -\theta_i <0.
 	\end{cases}
 \end{equation}
 If $J_{ij}=J < 0$ for all $i,j$, 
 the previous equation reads
 \begin{equation}
 	s_i(t+1) = 
 	\begin{cases}
 		+1 & \text{if} \quad  m(t)<\frac{\theta_i}{JN}, \\
 		-1 & \text{if} \quad  m(t)>\frac{\theta_i}{JN}.
 	\end{cases}
 \end{equation}
Since $m(t) \in \{-1,-1+\frac{2}{N}, \dots,\- +1-\frac{2}{N}, +1\}$, all the sites such that $\frac{\theta_i}{JN}<-1$ ($\frac{\theta_i}{JN}>1$) are frozen in the $-1$ ($+1$) state.
Let us suppose that this is valid for all spins except the $i$-th one, that can therefore flip. Only two possible magnetizations are possible in this case: one with $s_i=+1$, which we will call $S_2$, and one with $s_i = -1$, which we will call $S_1<S_2$;  if $S_1<\theta_i/JN<S_2$,  $s_i$ oscillates between the up and down states. A very simple case is shown in Fig.\ \ref{fig:oscillazione}: all the supports satisfy the condition $\theta_j/JN<S_1$ except the $i$-th one, that satisfies the condition $S_1<\theta_i/JN<S_2$. This means that $S_1=-1$ and $S_2=-1+2/N$. When the magnetization of the system is $-1$, the $i$-th spin flips, so the magnetization becomes $-1+2/N$, then the spin flips again bringing the magnetization back to the previous value. This oscillation goes on indefinitely. 
 
If $N-2$ spins satisfy the condition $\theta_j/JN<S_1$ and $2$ of them satisfy the condition $S_1<\theta_i<S_2$ a similar behavior occurs, but $S_2-S_1=4/N$. Of course these oscillations take place only if the interactions are negative. Incidentally, this explains why no limit cycle is observed in Fig.\ \ref{fig:magnetizzazione_asintotica_conflittuale}.
 
Summarizing, the above discussion implies that when the above-mentioned conditions hold, two different asymptotic magnetizations appear, one for even and the other one for odd times. Both tend to vanish when the standard deviations $\sigma_J,\sigma_{\xi}\rightarrow \infty$. We have thus unveiled the appearance in the system of an attractive limit cycle of period 2 \cite{limitcycle,limitcycle1}. 
As will be clear from the following figures, the size of the cycle depends on the supports and shrinks for large values of the noise and/or the couplings. This yields important information about the typical values of $\sigma_{\xi}$ and $\sigma_{J}$ such that the evolving system does not feel any difference between even and odd times.

\begin{figure}
 	\centering
 	\includegraphics[width=\columnwidth]{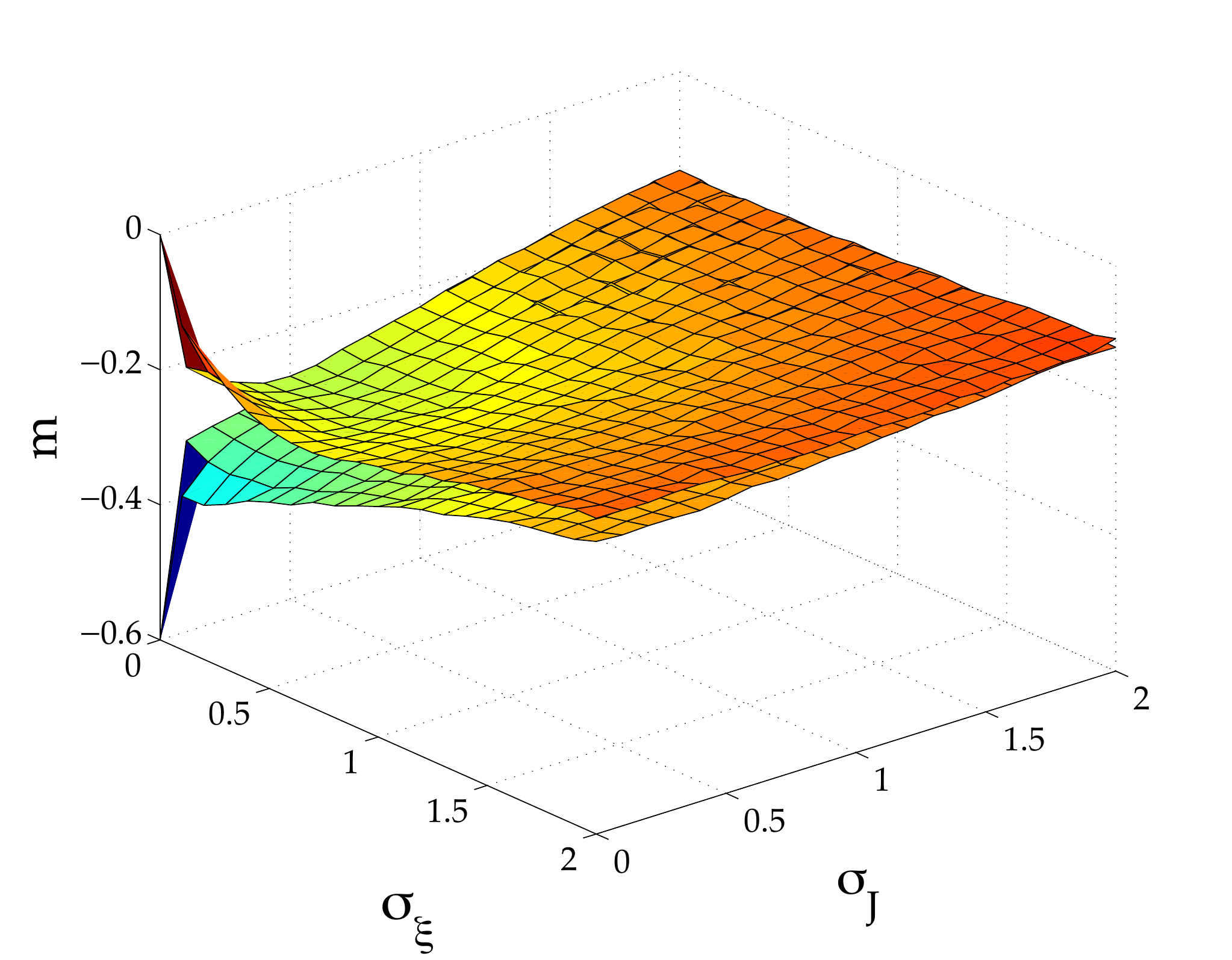}
 	\includegraphics[width=\columnwidth]{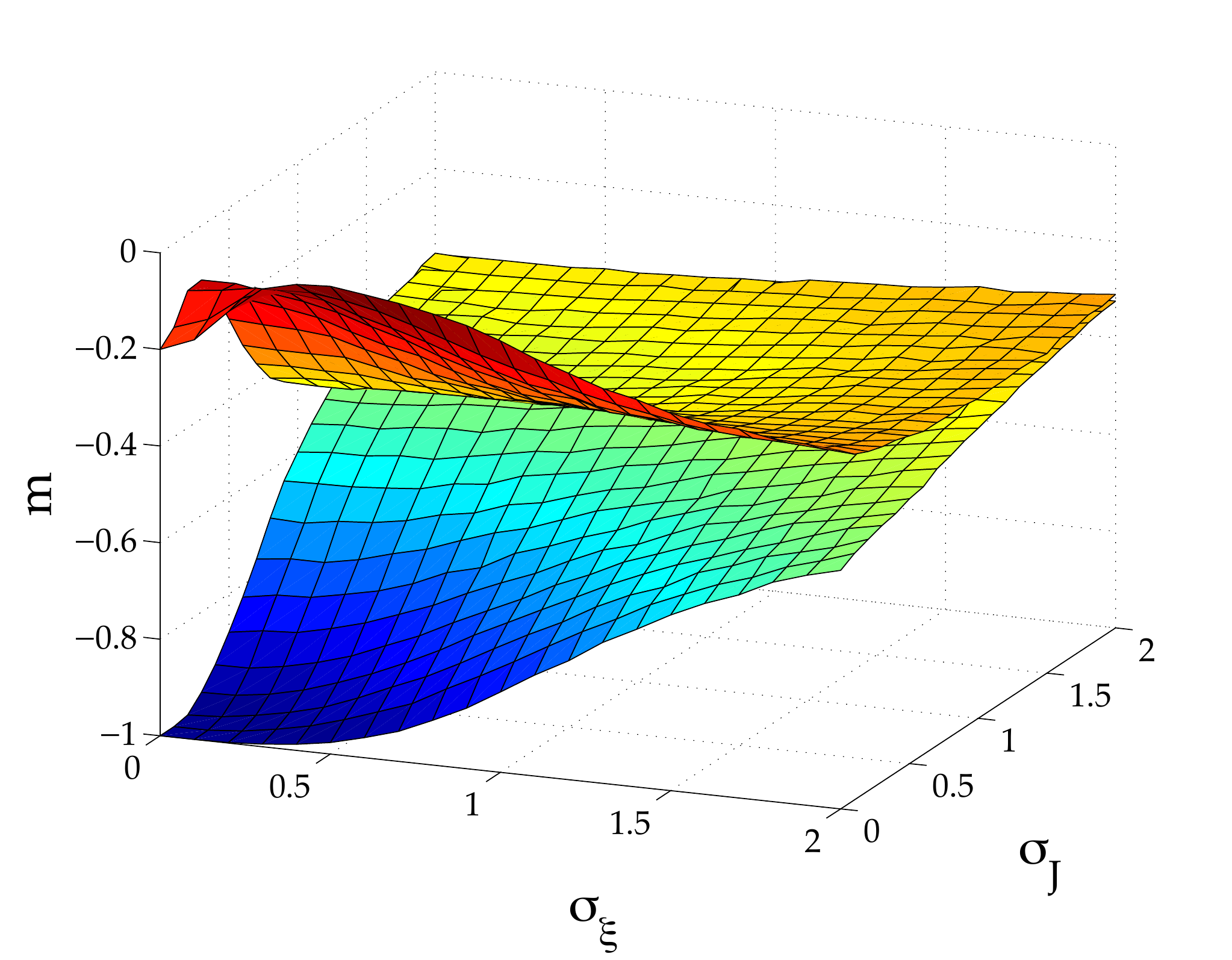}
 	\caption{(Color online) Asymptotic magnetization vs $\sigma_J$ and $\sigma_{\xi}$. The couplings are chosen according to \eqref{eq:funzione_distribuzione}, while the supports are extracted from a Gaussian distribution. Upper panel: $\mu_\theta=0.5999$ and $\sigma_\theta=0.4684$; lower panel: $\mu_\theta=1.6602$ and $\sigma_\theta=0.3308$. A limit cycle appears: the two leaves represent the asymptotic magnetization for odd and even times.}
\label{fig:theta_fixed}	
 \end{figure}

Figure\ \ref{fig:theta_fixed} displays the behavior of the asymptotic magnetization for even and odd times and for two different realizations of the supports. In the upper panel $\vec{\theta}=\;$(0.8636, 1.1562,    0.5200,    0.1114,    0.1399,    0.8885,    1.2400,    0.2851,    0.8793,   -0.0849) ($\mu_\theta=0.5999$ and $\sigma_\theta=0.4684$), while in the lower panel $\vec{\theta}=\;$(1.1576, 1.9706, 1.9572, 1.4854, 1.8003, 1.1419, 1.4218, 1.9157, 1.7922, 1.9595) ($\mu_\theta=1.6602$ and $\sigma_\theta=0.3308$). The values are averaged over several realizations of the couplings, distributed according to (\ref{eq:funzione_distribuzione}). The main difference is the behavior near the origin: clearly, larger supports entail more inertia in the spin dynamics and a slower convergence to the asymptotic value.

 \begin{figure}[t]
 	\centering
 	\includegraphics[width=\columnwidth]{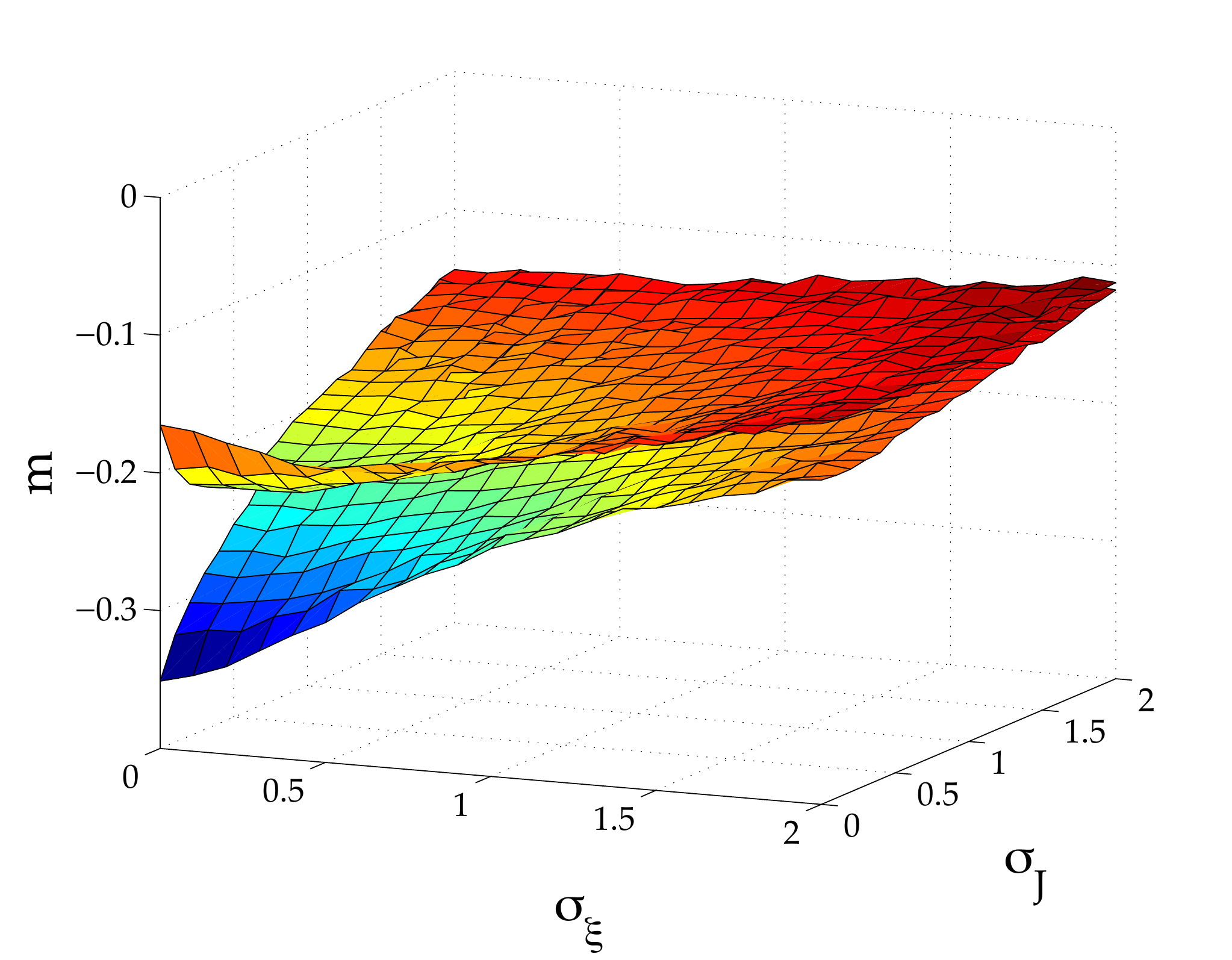}
 	\includegraphics[width=\columnwidth]{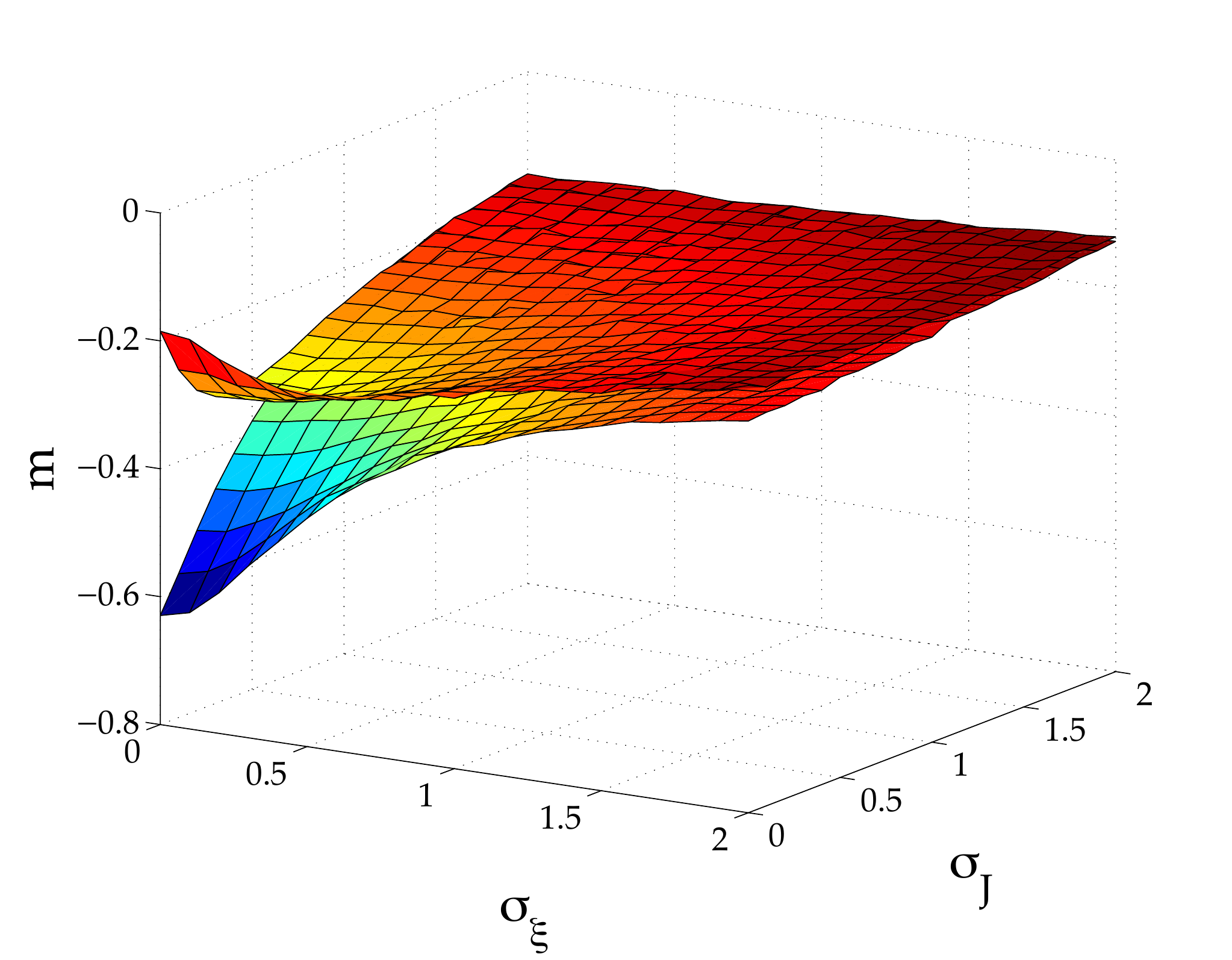}
 	\caption{(Color online) Asymptotic magnetization vs $\sigma_J$ and $\sigma_{\xi}$. The couplings are chosen according to \eqref{eq:funzione_distribuzione}, while the supports are quenched random variables extracted from a Gaussian distribution with $\mu_\theta= 0.5$. Upper panel $\sigma_\theta = 0.5$; bottom panel $\sigma_\theta = 1$. Like in Fig.\ 
\ref{fig:theta_fixed}, a limit cycle appears: the two leaves represent the asymptotic magnetization for odd and even times.}
\label{fig:theta_quenched}
 \end{figure}
\begin{figure}[t]
 	\centering
 	\includegraphics[width=\columnwidth]{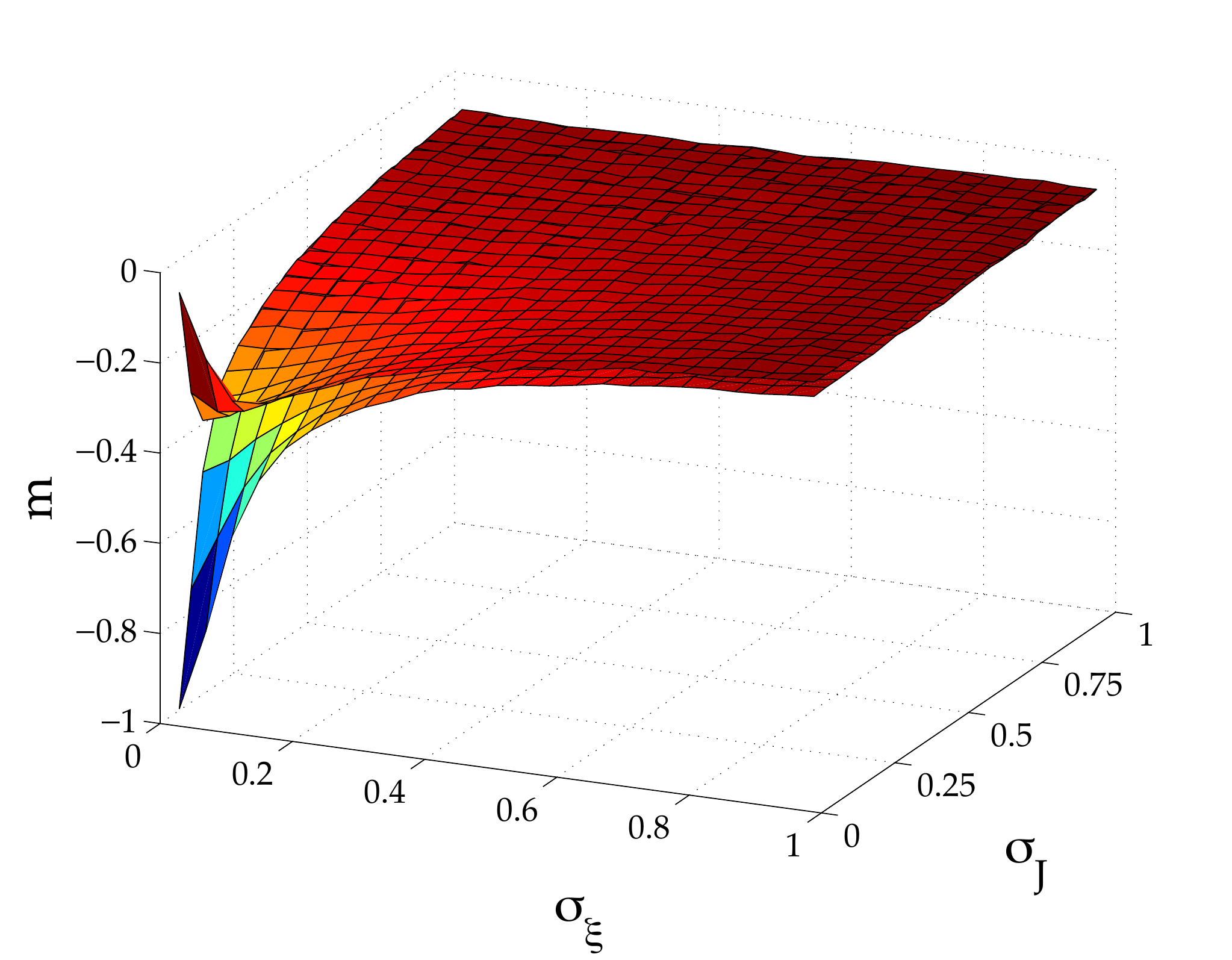}
 	\includegraphics[width=\columnwidth]{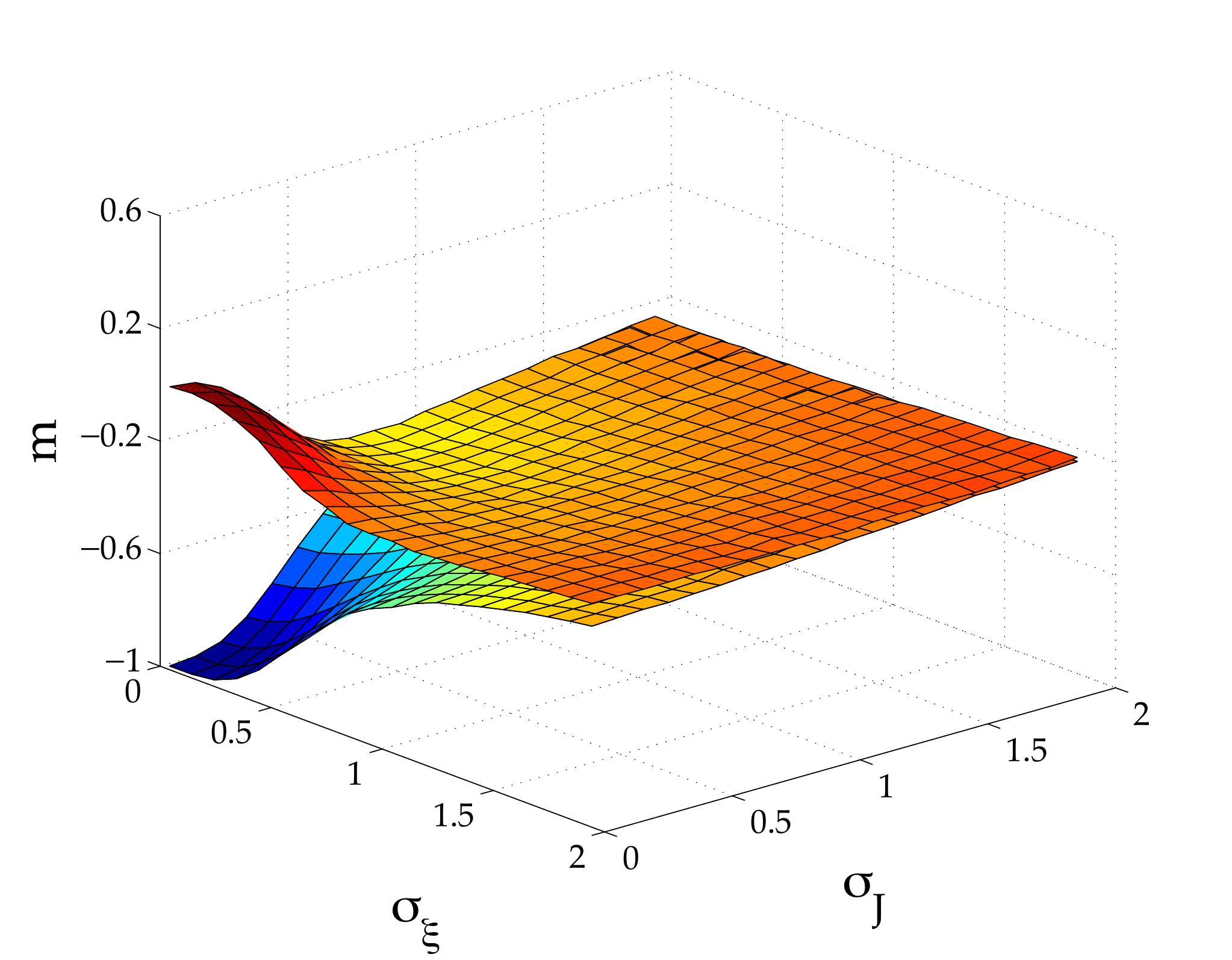} 
 	\caption{ (Color online) Asymptotic magnetization vs $\sigma_J$ and $\sigma_{\xi}$. The couplings are chosen according to \eqref{eq:funzione_distribuzione}, while the supports are all equal: in the upper panel, $\theta_i=\theta=0.1, \; \forall i$, while in the lower panel  $\theta_i=\theta=0.8, \; \forall i$. Again, a limit cycle appears, the two leaves representing the asymptotic magnetization for odd and even times. The manifolds appear to be more regular than in Figs.\ \ref{fig:theta_fixed} and \ref{fig:theta_quenched}.}
\label{fig:theta_equal}
 \end{figure}

In Fig.\ \ref{fig:theta_equal} all supports are equal ($\sigma_\theta=0$). In the left panel $\theta_i=\theta= 0.1, \forall i$, while in the right panel $\theta_i= \theta=0.8, \forall i $. As is to be expected, the manifolds appear to be more regular, in particular for larger $\theta$.

In Fig.\ \ref{fig:theta_quenched} we focus on the role of the standard deviation of the distribution of the supports. 
Again, the supports are quenched random variables, drawn from a Gaussian distribution with $\mu_{\theta} = 0.5$. In the upper panel $\sigma_{\theta} = 0.5$ and in the lower panel $\sigma_{\theta} = 1$. The manifolds are similar and differ only for their behavior near the origin: as stressed before, in that region the behavior of the magnetization strongly depends on the supports $\theta_i$'s (see Fig.\ \ref{fig:confront_high_low_theta}). The two manifolds merge more easily for larger $\sigma_\theta$. 

\begin{figure}
  \centering
    {\includegraphics[width=0.9\columnwidth]{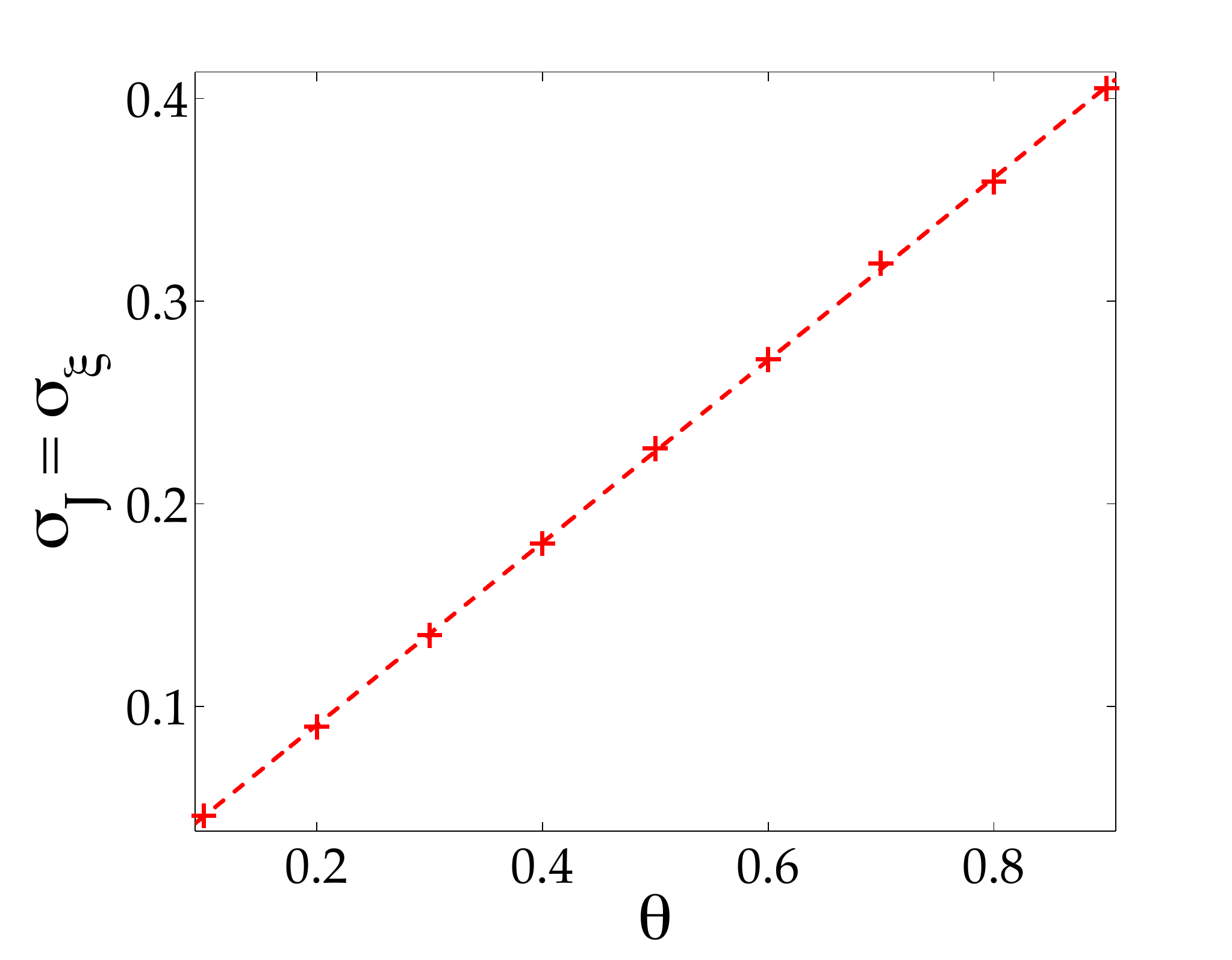}}
  \hspace{2mm}
    {\includegraphics[width=0.9\columnwidth]{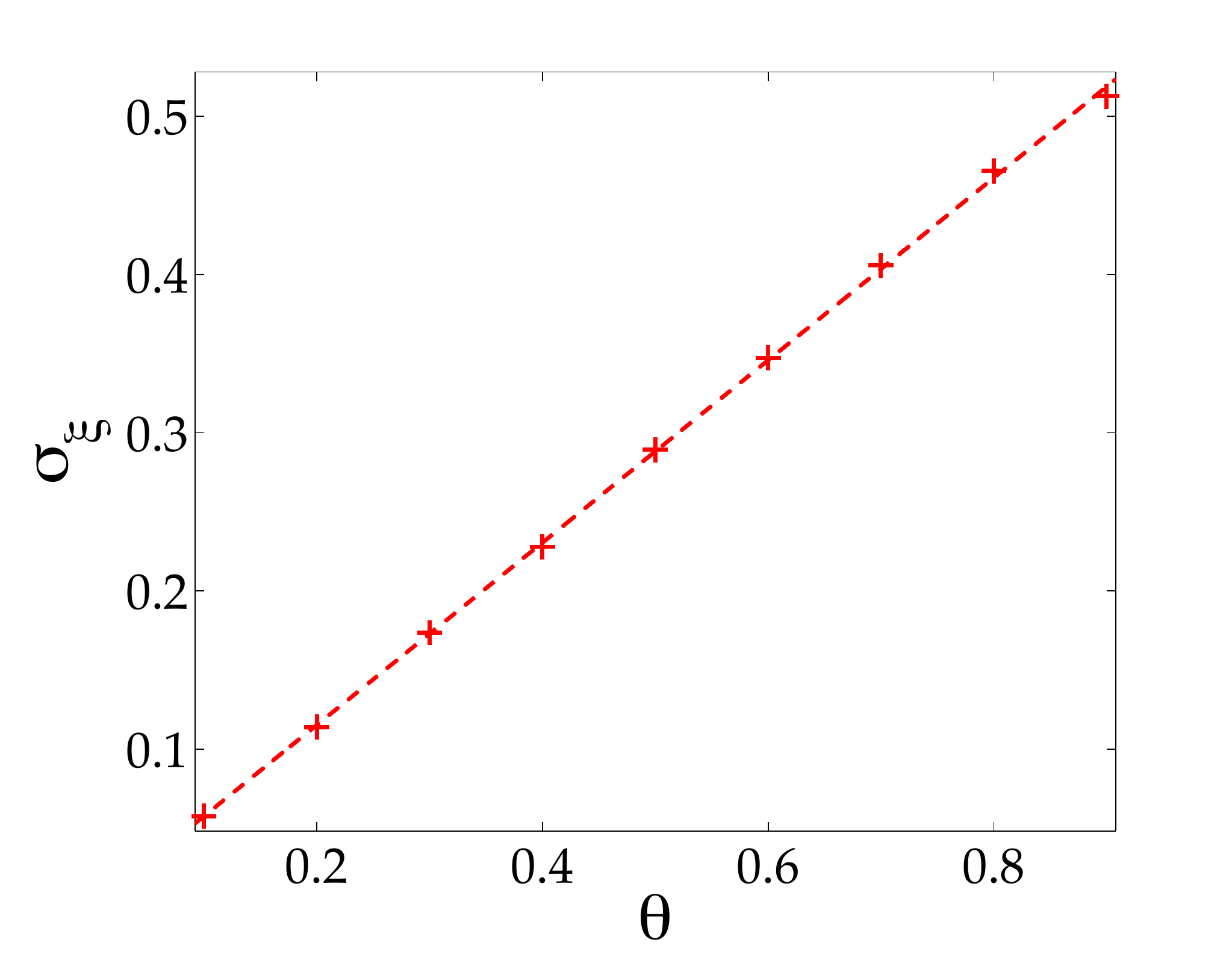}}
 \hspace{2mm}
    {\includegraphics[width=0.9\columnwidth]{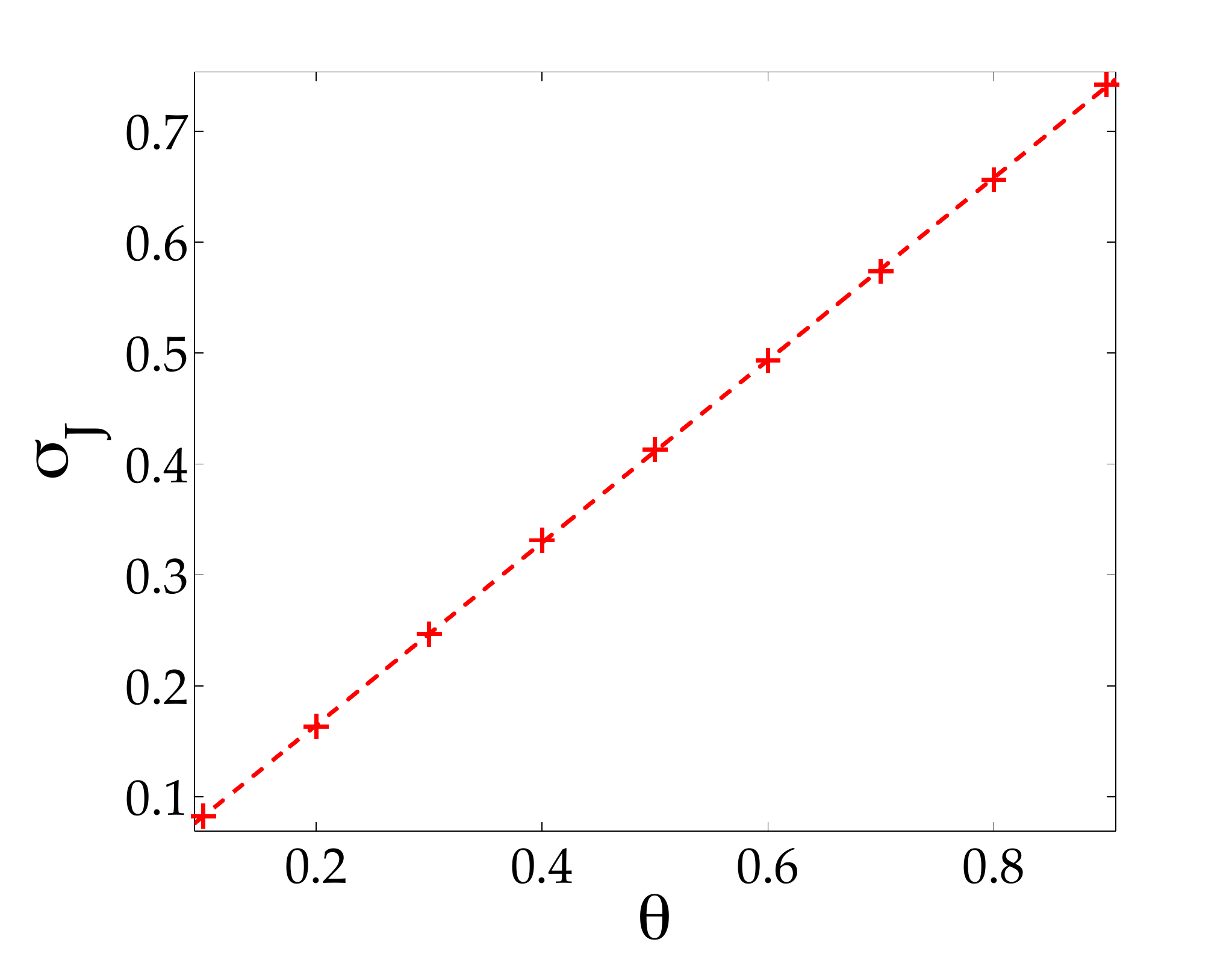}}
  \caption{(Color online) Values of the standard deviations at which the radius of the limit cycle (difference between magnetization at odd and even times) is $1/4$. We look at three different directions: $\sigma_J=\sigma_{\xi}$ (upper panel); $\sigma_{\xi}=0$, (central panel); and $\sigma_J=0$ (lower panel). The linear fits $\sigma(\theta)=a\theta + b$ have slopes $a=0.45, 0.58, 0.82$, respectively. Their correlation coefficients  $\rho$ are bounded by $\rho^2\geq0.9996$ and the coefficient $b$ is compatible with 0, namely $|b|\leq 8.6 \times 10^{-4}$.
}
\label{fig:linear_plots}
\end{figure}

Interestingly, a curious feature emerges in the case of uniform supports: Figure \ref {fig:linear_plots} shows the value of the standard deviation at which the distance between even and odd magnetizations (radius of the limit cycle) is $1/4$.
In the upper panel $\sigma_{\xi}=\sigma_J$; in the central panel $\sigma_{\xi} = 0$; in the lower panel $\sigma_J = 0$. A linear fit yields excellent accord with the numerical data. We offer no explanation for this linear behavior.

 \section{Conclusions}
 \label{sec:conclusion}

We analyzed loss risks of banks by borrowing techniques of classical statistical mechanics.
Losses were divided into $N$ categories, each of which corresponds to a channel, schematized as a dichotomic variable (spin). This is paramount to assuming that one is only interested in the occurrence of a loss and not in its amount. The basic idea is that a loss occurring in a channel can stimulate (inhibit) losses in other channels: this is mathematically described by assuming that loss events interact with each other by means of positive or negative couplings constants
\cite{kuhn,Leippold:2005,Clemente:2004,anand}. We assumed that the interactions are random, as in a glassy system. 

The time evolution of this model is governed by Eq.\ (\ref{evolution_equation}) and we studied the asymptotic behavior of the system for interesting values of the parameters. Our study brought to light some interesting features of the model. First of all, the system has a short memory: it quickly forgets its initial spin configurations and the magnetization quickly converges towards its asymptotic value, which is determined by the supports and the parameters of the distribution function of noise and disorder. One should remember that the magnetization represents the average number of channels in which losses occur. 

Interesting features appear when the parameters are rescaled as in Eq.\ (\ref{eq:funzione_distribuzione}) and the supports are positive (we remind that a positive support means that a bank gives ``care" for a source of potential losses in a given channel).
For small values of the standard deviations of noise and disorder the system displays an oscillatory behavior (between even and odd time steps). This is a limit cycle \cite{limitcycle,limitcycle1}.
For larger values of these standard deviations, the system reaches a unique asymptotic value. This yields a scale on which the behavior of the system can be regarded as stable. In general, this scale depends on the parameters of the model. However, when the supports are identical, this scale \emph{linearly} depends on the (common) support. We have offered no explanation for this curious numerical observation.

In recent years, the scientific interest for economical issues has increased, thanks also to the contribution of advanced physical and mathematical techniques, such as  stochastic processes, chaos theory and the theory of disordered systems \cite{mandelbrot,MantegnaStanley}. This work is an effort towards the modelling of the operational risk of banks activities. Much additional work must be done, in particular in order to understand whether a glassy phase transition appears. An additional point of interest, also in view of potential applications, would be to work towards the optimization of some crucial parameters of the model, such as the supports. This would enable banks to understand in which direction their efforts should be directed in order to minimize losses.


\begin{thebibliography}{99}
 
\bibitem{GK}
N. Goldenfeld and L. P. Kadanoff, Science \textbf{284}, 87 (1999).

\bibitem{parisi}
M. Mezard, G. Parisi and M. A. Virasoro, \textit{Spin Glass Theory and
Beyond} (World Scientific, Singapore, 1987).

\bibitem{dotsenko}
V. Dotsenko, \textit{An Introduction to the Theory of Spin Glasses and Neural Networks} (World Scientific, Singapore, 1995).

\bibitem{mandelbrot}
B. Mandelbrot, \textit{Fractals and Scaling in Finance: Discontinuity, Concentration, Risk} (Springer-Verlag, New York, 1997). 
	
\bibitem{MantegnaStanley}
 R. N. Mantegna and H. E. Stanley,
 \textit{An Introduction to Econophysics: Correlations and Complexity in Finance}
(Cambridge University Press, Cambridge UK, 2000).

\bibitem{cruz}
M. G. Cruz, \textit{Modeling, Measuring and Hedging Operational Risk} (Wiley, Chichester, 2002).

\bibitem{king}
J. L. King, \textit{Operational Risk: Measurement and Modelling} (Wiley, Chichester, 2002).

\bibitem{basel}
Basel Committee on Banking Supervision, \textit{International Convergence of Capital Measurement and Capital Ctandards} (Bank for International Settlements Press \& Communications, 2005).

\bibitem{embrechts}
A. J. McNeil, R. Frey and P. Embrechts, \textit{Quantitative Risk Management: Concepts, Techniques and Tools} (Princeton University Press, Princeton, 2005).

\bibitem{kuhn}
R. Kh\"{u}n and P. Neu,
Physica A \textbf{322}, 650 (2003).

\bibitem{Leippold:2005}
M. Leippold and P. Vanini,
J. Risk \textbf{8}, 59 (2005).

\bibitem{Clemente:2004}
A. D. Clemente and C. Romani,
``A Copula-Extreme Value Theory Approach for Modelling Operational 
Risk", Chapter in {\it Operational Risk Modelling and Analysis: Theory and Practice}
(Risk Books, London, 2004).

\bibitem{anand}
R. Kh\"{u}n and K. Anand,
Phys. Rev. E \textbf{75}, 016111 (2007).

\bibitem{limitcycle}
S. H. Strogatz, \textit{Nonlinear Dynamics and Chaos} (Addison Wesley, Reading, Massachusetts, 1994).

\bibitem{limitcycle1}
J. Guckenheimer and P. Holmes, \textit{Nonlinear Oscillations, Dynamical Systems and Bifurcations of Vector Fields} (Springer-Verlag, Berlin, 1983)



\end{thebibliography}
\end{document}